\documentclass{vldb}
\usepackage{defpaper2e}

\usepackage{graphicx}
\usepackage{color}
\usepackage{amsmath,epsfig}
\usepackage{epstopdf}
\usepackage{multirow} %\usepackage{defpaper2e}
\usepackage{amssymb,amsmath}
\usepackage{hhline}
\usepackage{url}
\usepackage{multirow}

\newcommand{\algorithmicbreak}{\textbf{break}}
\newcommand{\BREAK}{\STATE \algorithmicbreak}
\newenvironment{ditemize}{%   
\begin{list}{{\bf $\bullet$}}{%
\setlength{\itemsep}{0pt}\setlength{\rightmargin}{0pt}%
\setlength{\leftmargin}{1.2em}\setlength{\parsep}{0pt}}}{
\end{list}}

\title{Frequency-hiding Dependency-preserving Encryption for Outsourced Databases}

\numberofauthors{1} %  in this sample file, there are a *total*
\author{
\alignauthor
Boxiang Dong, Wendy Wang\\
       \affaddr{Department of Computer Science}\\
       \affaddr{Stevens Institute of Technology}\\
       \affaddr{Hoboken, NJ}\\
       \email{bdong, hwang4@stevens.edu}
}

\begin{document}

\maketitle

\begin{abstract} The cloud paradigm enables users to outsource their data to computationally powerful third-party service providers for data management. Many data management tasks rely on the data dependencies in the outsourced data. This raises an important issue of how the data owner can protect the sensitive information in the outsourced data while preserving the data dependencies. In this paper, we consider {\em functional dependency} ($FD$), an important type of data dependency. 
We design a $FD$-preserving encryption scheme, named $F^2$, that enables the service provider to discover the $FDs$ from the encrypted dataset. We consider the frequency analysis attack, and show that the $F^2$ encryption scheme can defend against the attack under Kerckhoff's principle with provable guarantee. 
  Our empirical study demonstrates the efficiency and effectiveness of $F^2$. 
\end{abstract}

\vspace{-.05in}
\section{Introduction}

With the fast growing of data volume, the sheer size of today's data sets is increasingly crossing the petabyte barrier, which far exceeds the capacity of an average business computer. In-house solutions may be expensive due to the purchase of software, hardware, and staffing cost of in-house resources to administer the software. 
%It is not uncommon for system costs to exceed \$1 million \cite{tomos}. 
Lately, due to the advent of cloud computing and its model for IT services based on the Internet and big data centers, a new model has emerged to give companies a cheaper alternative to in-house solutions: the users outsource their data management needs to a third-party service provider. 
%A client pays a per-user fee for a predefined period of time to access the solution. 
Outsourcing of data and computing services is becoming commonplace and essential.  

%This emerges the {\em database-as-a-service ($DaS$)} paradigm. 
%Among various outsourcing services that are provided, database-as-a-service ($DaS$) is an architectural and operational approach that enables IT providers to deliver database functionality as a service to one or more consumers \cite{oracledaas}. 
%$DaS$ provides various services such as data storage and access, database architecture design and implementation,  as well as data query evaluation and optimization. 
%$DaS$ provides various data services such as data storage \cite{das2009elastras,kraska2009consistency,curino2011relational} and query evaluation \cite{popa2012cryptdb,hacigumucs2002executing}. 

Many outsourced data management applications rely on data dependencies in the outsourced datasets. 
A typical type of data dependency is {\em functional dependency} (FD). 
%\cite{huhtala1999tane} and conditional functional dependency (CFD) \cite{chiang2008discovering,golab2008generating,DBLP:journals/tkde/FanGLX11}. In this paper, we consider $FDs$ as the main ICs. 
Informally, a $FD: A\rightarrow B$ constraint indicates that an attribute set $A$ uniquely determines an attribute set $B$. For example, the FD {\em Zipcode}$\rightarrow${\em City} indicates that all tuples of the same {\em Zipcode} values always have the same {\em City} values. FDs serve a wide range of data applications, for example, improving schema quality through normalization \cite{batini1986comparative,chiticariu2007semi}, and improving data quality in data cleaning \cite{DBLP:journals/tkde/FanGLX11}.  
Therefore, to support these applications in the outsourcing paradigm, it is vital that FDs are well preserved in the outsourced datasets.

\nop{
\begin{figure}[t!]
\begin{center}
\begin{small}
\begin{tabular}{cc}
\begin{tabular}{|l|l|l|}
\hline
Name&Zipcode&City\\
\hline
Alice & 10001 & NYC\\\hline
Bob & 10001 & NYC \\\hline
Carol & 10002 & NYC \\\hline
Dan & 07030 & Hoboken\\\hline
\end{tabular}
&
\begin{tabular}{|l|l|l|}
\hline
Name&Zipcode&City\\
\hline
P98TR & VX32W & 1YBTE\\\hline
A81Q2 & VX32W & 1YBTE \\\hline
6REWC & SOFU7 & 1YBTE \\\hline
O84WQ & MRE2Q & 75EQZ\\\hline
\end{tabular}
\\
(a) Base table $D$ &
(b) Encrypted table $\hat{D}$ by 
\\
 (FD: Zipcode$\rightarrow$ City) &
applying deterministic substitution on $D$ 
\end{tabular}
\vspace{-0.1in}
\caption{\label{table:first} An example showing that deterministic encryption preserves FDs (but not secure)}
\end{small}
\end{center}
\vspace{-0.3in}
%\end{figure*}
\end{figure}
}

\begin{figure}[t!]
\begin{center}
\begin{small}
\begin{tabular}{cc}
\begin{tabular}{|l|l|l|l|}
\hline
ID&A&B&C\\
\hline
$t_1$&$a_1$ & $b_1$ & $c_1$\\\hline
$t_2$&$a_1$ & $b_1$ & $c_2$\\\hline
$t_3$&$a_1$ & $b_1$ & $c_3$\\\hline
$t_4$&$a_1$ & $b_1$ & $c_1$\\\hline
\end{tabular}
&
\begin{tabular}{|l|l|l|l|}
\hline
ID&A&B&C\\
\hline
$t_1$&$\hat{a}_1$ & $\hat{b}_1$ & $\hat{c}_1$\\\hline
$t_2$&$\hat{a}_1$ & $\hat{b}_1$ & $\hat{c}_2$\\\hline
$t_3$&$\hat{a}_1$ & $\hat{b}_1$ & $\hat{c}_3$\\\hline
$t_4$&$\hat{a}_1$ & $\hat{b}_1$ & $\hat{c}_1$\\\hline
\end{tabular}
\\
(a) Base table $D$ 
&
(b) $\hat{D}_1$ encrypted 
\\
($FD: A\rightarrow B$) 
&
by deterministic encryption
\\
&
%($FD: A\rightarrow B$ in $\hat{D}_1$) 
(not frequency-hiding)
\\
\begin{tabular}{|l|l|l|l|}
\hline
ID&A&B&C\\
\hline
$t_1$&$\hat{a}_1^1$ & $\hat{b}_1^1$ & $\hat{c}_1^1$\\\hline
$t_2$&$\hat{a}_1^1$ & $\hat{b}_1^2$ & $\hat{c}_2^1$\\\hline
$t_3$&$\hat{a}_1^2$ & $\hat{b}_1^1$ & $\hat{c}_3^1$\\\hline
$t_4$&$\hat{a}_1^2$ & $\hat{b}_1^2$ & $\hat{c}_1^2$\\\hline
\end{tabular}
&
\begin{tabular}{|l|l|l|l|}
\hline
ID&A&B&C\\
\hline
$t_1$&$\hat{a}_1^1$ & $\hat{b}_1^1$ & $\hat{c}_1^1$\\\hline
$t_2$&$\hat{a}_1^1$ & $\hat{b}_1^1$ & $\hat{c}_2^1$\\\hline
$t_3$&$\hat{a}_1^2$ & $\hat{b}_1^2$ & $\hat{c}_3^1$\\\hline
$t_4$&$\hat{a}_1^2$ & $\hat{b}_1^2$ & $\hat{c}_1^2$\\\hline
\end{tabular}
\\
(c) $\hat{D}_2$ encrypted by 
&
(d) $\hat{D}_3$ encrypted by 
\\
probabilistic encryption on 
&
probabilistic encryption 
\\
$A$ and $B$ individually 
&
on attribute set $\{A, B\}$
\\
(not FD-preserving)
&
(Frequency-hiding \& FD-preserving)
\end{tabular}
\vspace{-0.1in}
\caption{\label{table:release} An example of various encryption schemes}
\end{small}
\end{center}
\vspace{-0.3in}
\end{figure}

Outsourcing data to a potentially untrusted third-party service provider (server) raises several security issues. One of the issues is to protect the sensitive information in the outsourced data. The data confidentiality problem is traditionally addressed by means of encryption \cite{davida1981database}. 
Our goal is to design efficient {\em FD-preserving} data encryption methods, so that the FDs in the original dataset still hold in the encrypted dataset. 
A naive method is to apply a simple deterministic encryption scheme (i.e. the same plaintext values are always encrypted as the same ciphertext for a given key) on the attributes with FDs. For instance, consider the base table $D$ in Figure \ref{table:release} (a). Assume it has a FD: $F$: $A \rightarrow B$. Figure \ref{table:release} (b) shows the encrypted dataset $\hat{D}_1$ by applying a deterministic encryption scheme on individual values of the attributes in $D$. Apparently $F$ is preserved in $\hat{D}_1$. However, this naive method has drawbacks.
One of the main drawbacks is that the deterministic encryption scheme is vulnerable against the frequency analysis attack, as the encryption preserves the frequency distribution. The attacker can easily map the ciphertext (e.g., $\hat{a}_{1}$) to the plaintext values (e.g., $a_1$) based on their frequency. 
\nop{ 
First, this method assumes that $FD$s are available before encryption, which may not hold in practice.
the assumption that $FDs$ were determined in advance may not hold in practice \cite{yu2008xml}. Indeed, finding $FDs$ needs intensive computational efforts \cite{huhtala1999tane}. As shown in a recent study \cite{papenbrock2015functional}, for a dataset with 100 columns and $300k$ rows, finding FDs will take days or longer. 
%Even for smaller datasets (e.g., 40 columns and 100,000 rows), the algorithm needs 256 GB main memory or more.  
Under such circumstances, the data owner may lack computational resources and/or knowledge to discover FDs by herself before outsourcing her data to the service provider. 
}
%Another weakness is that it assumes that $FD$s are available to the client, which may not hold in practice. Finding $FDs$ needs intensive computational efforts \cite{huhtala1999tane}. The client may be lack of computational resources and/or knowledge to discover FDs by herself.
%As shown previously, the frequency-based adversary knowledge can be easily obtained in real-world applications \cite{schnell2009privacy}.  

A straightforward solution to defend against the frequency analysis attack is instead of the deterministic encryption schemes, using the probabilistic encryption schemes (i.e., the same plaintext values are encrypted as different ciphertexts) to hide frequency. Though the probabilistic encryption can provide provable guarantee of semantic security \cite{goldwasser1984probabilistic}, it may destroy the FDs in the data.  As an example, consider the table $D$ shown in Figure \ref{table:release} (a) again. Figure \ref{table:release} (c) shows an instance $\hat{D}_2$ by applying the probabilistic encryption scheme at the attributes $A$, $B$, and $C$ individually. We use the notation $\hat{a}_{i}^{j}$ ($\hat{b}_{i}^{j}$, $\hat{c}_{i}^{j}$, resp.) as a ciphertext value of $a_i$ ($b_i$, $c_i$, resp.) by the probabilistic encryption scheme. The four occurrences of the plaintext value $a_1$ are encrypted as two $\hat{a}_{1}^{1}$s (tuple $t_1$ and $t_2$) and two $\hat{a}_{1}^{2}$s (tuple $t_3$ and $t_4$), while the four occurrences of the plaintext value $b_1$ are encrypted as two $\hat{b}_{1}^{1}$s  (tuple $t_1$ and $t_3$) and two $\hat{b}_{1}^{2}$s  (tuple $t_2$ and $t_4$). Now the frequency distribution of the plaintext and ciphertext values is not close. However, the FD $A\rightarrow B$ does not hold on $\hat{D}_2$ anymore, as the same (encrypted) value $\hat{a}_1$ of attribute $A$ is not associated with the same cipher value of attribute $B$. 
This example shows that the probabilistic encryption scheme, if it is applied on individual attributes, cannot preserve the FDs. 
This raises the challenge of finding the appropriate attribute set to be encrypted by the probabilistic encryption scheme. 
It is true that applying the probabilistic encryption scheme on FD attributes as a unit (e.g., on \{A, B\} as shown in Figure \ref{table:release} (d)) can preserve FDs. However, finding $FDs$ needs intensive computational efforts \cite{huhtala1999tane}. 
%As shown in a recent study \cite{papenbrock2015functional}, for a dataset with 100 columns and $300k$ rows, finding FDs will take days or longer. Under such circumstances, 
The data owner may lack of computational resources and/or knowledge to discover FDs by herself. Therefore, we assume that FDs are not available for the encryption.

{\bf Contributions.} In this paper, we design $F^2$, a \underline{f}requency-hiding, \underline{F}D-preserving encryption scheme based on probabilistic encryption.
%We focus on the relational data model in which data is stored in flat, two-dimensional tables. 
%First, we design efficient and {\em FD-preserving} encryption methods by which $FDs$ in the original dataset still hold in the encrypted data. This enables the server to discover $FDs$ from the outsourced (encrypted) data, and use $FDs$ for schema refinement.
We consider the attacker who may possess the frequency distribution of the original dataset as well as the details of the encryption scheme as the adversary knowledge. 
 %Therefore, cell-based encryption may not be secure against the attack based on the frequency analysis of the encrypted data. Thus, second,  
%To our best knowledge, our work is the first that considers how to preserve useful FDs with provable security guarantees in the outsourcing paradigm. 
We make the following contributions. 
 
(i) $F^2$ allows the data owner to encrypt the data without awareness of any FD in the original dataset. To defend against the frequency analysis attack on the encrypted data, we apply the {\em probabilistic encryption} in a way that the frequency distribution of the ciphertext values is always flattened regardless of the original frequency distribution. This makes $F^2$ robust against the frequency analysis attack. To preserve FDs, we discover {\em maximal attribute sets} ($MASs$) on which $FDs$ {\em potentially} exist. Finding $MASs$ is much cheaper than finding $FDs$, which enables the data owner to afford its computational cost. The probabilistic encryption is applied at the granularity of $MAS$s. 
%The Homophonic substitution encoding is applied on $MASs$. 

(ii) We consider both cases of one single $MAS$ and multiple $MASs$, and design efficient encryption algorithms for each case. The challenge is that when there are multiple $MASs$, there may exist {\em conflicts} if the encryption is applied on each $MAS$ independently. We provide {\em conflict resolution} solution, and show that the incurred amounts of overhead by conflict resolution is bounded by the size of the dataset as well as the number of overlapping $MAS$ pairs.   

(iii) Applying probabilistic encryption may introduce {\em false positive} FDs that do not exist in the original dataset but in the encrypted data. We eliminate such false positive FDs by adding a small amount of artificial records. We prove that the number of such artificial records is independent of the size of the outsourced dataset. 

(iv) We formally define the frequency analysis attack, and the {\em $\alpha$-security} model to defend against the frequency analysis attack. We show that $F^2$ can provide provable security guarantee against the frequency analysis attack, even under the Kerckhoffs's principle \cite{shannon1949communication} (i.e., the encryption scheme is secure even if everything about the encryption scheme, except the key, is public knowledge). 
%two attacks: (1) the {\bf Freq+$\hat{D}$} attack that exploits the frequency distribution of both original and encrypted data, and (2) the {\bf Freq+$F^2$} attack that is launched based on the details of the $F^2$ algorithm.  We analyze the security guarantee of $F^2$ against these two attacks.

(v) We evaluate the performance of $F^2$ on both real and synthetic datasets. The experiment results show that $F^2$ can encrypt large datasets efficiently with high security guarantee. For instance, it takes around 1,500 seconds to encrypt  a benchmark dataset of size 1GB with high security guarantee. Furthermore, $F^2$ is much faster than FD discovery. For instance, it takes around 1,700 seconds to discover FDs from a dataset of 64KB records, while only 8 seconds to encrypt the dataset by $F^2$. 

\noindent{\bf Applications} This  is the first algorithm that preserves useful FDs with provable security guarantees against the frequency analysis attack in the outsourcing paradigm. To see the relevance of this  achievement, 
 consider that FDs have been identified as a key component of several classes of database applications, including data schema refinement and normalization \cite{batini1986comparative, chiticariu2007semi, miller1993use}, data cleaning \cite{bohannon2007conditional, DBLP:journals/tkde/FanGLX11}, and schema-mapping and data exchange \cite{marnette2010scalable}. Therefore, we believe that the proposed FD-preserving algorithm represents a significant contribution towards the goal of reaching the full maturity of data outsourcing systems (e.g., database-as-a-service \cite{hacigumucs2002executing} and data-cleaning-as-a-service \cite{dong2014prada} systems). 

The rest of this paper is organized as follows. Section \ref{sc:pre} describes the preliminaries. Section \ref{sc:encryption} discusses the details of our encryption scheme. Section \ref{sc:security} presents the security analysis. Section \ref{sc:exp} evaluates the performance of our approach. Related work is introduced in Section \ref{sc:related}, and Section \ref{sc:conclu} concludes the paper. 

\vspace{-0.1in}
\section{Preliminaries}
\label{sc:pre}

\nop{
Table \ref{table:symbol} shows the notations that we use in this paper.

\begin{table}
\begin{center}
\begin{tabular}{ |c | c |}\hline
Symbol & Meaning \\\hline
 $D$ & original dataset \\\hline                      
 $D^S$ & the dataset outsourced to the server\\\hline                      
 $n$ & number of tuples in $D$  \\\hline  
 $m$ & number of attributes in $D$  \\\hline  
 $\cal P$ & the set of unique plain values in $D$ \\\hline 
 $\cal E$ & the set of unique cipher values in $D^S$ \\\hline 
\end{tabular}
\end{center}
\caption{Symbols}
\label{table:symbol}
\end{table}
}

\vspace{-0.05in}
%\subsection{Simple and Homophonic Substitution Schemes}

%Substitution ciphers are among the oldest encryption methods. Many variants of the substitution cipher have been developed, including mono-alphabetic systems that employ a fixed substitution. Examples of mono-alphabetic substitutions include the simple substitution and Homophonic substitution ciphers. For the simple substitution cipher, each unique plaintext is mapped to a specific ciphertext mapping via a one-to-one mapping.  The Homophonic Substitution ($HS$) cipher is a many-to-one substitution cipher in which multiple ciphertext symbols can map to one single plaintext symbol. 

\vspace{-0.05in}
\subsection{Outsourcing Setting}
We consider the outsourcing framework that contains two parties: the data owner and the service provider (server). 
The data owner has a private relational table $D$ that consists of $m$ attributes and $n$ records. 
We use $r[X]$ to specify the value of record $r$ on the attribute(s) $X$.  
To protect the private information in $D$, the data owner encrypts $D$ to $\hat{D}$, and sends $\hat{D}$ to the server. The server discovers the dependency constraints of $\hat{D}$. These constraints can be utilized for data cleaning \cite{DBLP:journals/tkde/FanGLX11} or data schema refinement \cite{batini1986comparative}. 
In principle, database encryption may be performed at various levels of granularity. 
However, the encryption at the coarse levels such as attribute and record levels may disable the dependency discovery at the server side. Therefore, in this paper, we consider encryption {\em at cell level} (i.e., each data cell is encrypted individually). 
Given the high complexity of FD discovery \cite{huhtala1999tane}, we assume that the data owner is {\em not} aware of any FD in $D$ before she encrypts $D$. 
%We use the AES algorithm as the encryption mechanism. 

\nop{
The server accepts SQL queries from the users. 
A user query $Q$ can be raised in two types: 
(1) {\em Point query}: $Q$ consists of the value-based constraint that is specified in the format $\cup_{1\leq i\leq m} A_i=a_i$, where $A_i$ is an attribute of $D$, and $a_i$ is a specific value of $A_i$.  
(2) {\em Range query}: $Q$ consists of the value-based constraint in the format $\cup_{1\leq i\leq m} A_i\in[a_i, b_i]$, where $A_i$ is an attribute of $D$, and $a_i, b_j$ are two specific values of $A_i$.  
For both types of queries, the server returns the records that satisfy the value-based constraints of $Q$;
In this paper, we consider AES algorithm as the encryption mechanism. The algorithm is applied at the cell level to support execution of the point queries on the encrypted data. The execution of range queries can be supported by order-preserving encryption or homomorphic encryption \cite{popa2012cryptdb}.
}

\subsection{Functional Dependency}
In this paper, we consider {\em functional dependency} (FD) as the data dependency that the server aims to discover.  Formally, given a relation $R$, there is a FD between a set of attributes $X$ and $Y$ in $R$ (denoted as $X\rightarrow Y$) if for any pair of records $r_1, r_2$ in $R$, if $r_1[X] =r_2[X]$, then $r_1[Y]=r_2[Y]$. %Otherwise we say $X\rightarrow Y$ is {\em negative} (denoted as $X\not\rightarrow Y$). 
We use $LHS(F)$ ($RHS(F)$, resp.) to denote the attributes at the left-hand (right-hand, resp.) side of the FD $F$. 
%We call $LHS(F)$ the determinant attributes of $F$, and $RHS(F)$ the dependent attributes of $F$. 
For any FD $F: X\rightarrow Y$ such that $Y\subseteq X$, $F$ is considered as {\em trivial}. In this paper, we only consider non-trivial $FDs$.
It is well known that for any FD $F: X \rightarrow Y$ such that $Y$ contains more than one attribute, $F$ can be decomposed to multiple functional dependency rules, each having a single attribute at the right-hand side. Therefore, for the following discussions, WLOG, we assume that the FD rules only contain one single attribute at the right-hand side. 
%We consider that the client is not aware of any $FD$ in her dataset. 
%We assume that the data is consistent with the $FDs$. 

%There exists monotone property among positive and negative FDs. In particular, the {\em upward closure} specifies that for any positive FD $F: X\rightarrow Y$, the FD $X' \rightarrow Y$ must also hold for any attribute set $X'$ such that $X\subseteq X'$. And the {\em downward closure} states that for any negative FD $X\not\rightarrow Y$, $X' \not\rightarrow Y$ must also hold for any attribute set $X'$ such that $X'\subseteq X$. %We also assume that the data is consistent with the $FDs$.

\vspace{-0.05in}
\subsection{Deterministic and Probabilistic Encryption}
Our encryption scheme $\Pi$ consists of the following three algorithms:
\vspace{-0.05in}
\begin{ditemize}
  \item {\bf $k \leftarrow KeyGen(\lambda)$} generates a key based on the input security parameter $\lambda$.
  \item {\bf $y \leftarrow Encrypt(x, k)$} encrypts the input value $x$ using key $k$ and outputs the ciphertext $y$.
  \item {\bf $x \leftarrow Decrypt(y, k)$} computes the plaintext $x$ based on the ciphertext $y$ and key $k$.
\end{ditemize}
\vspace{-0.05in}
Based on the relationship between plaintexts and ciphertexts, there are two types of encryption schemes: {\em deterministic} and {\em probabilistic} schemes. 
Given the same key and plaintext, a deterministic encryption scheme always generates the same ciphertext. 
One weakness of the deterministic ciphertexts is that it is vulnerable against the frequency analysis attack. 
%As the plaintexts and ciphertexts share the same frequency distribution, a matching based on frequency is sufficient to break the cipher. 
On the other hand, the probabilistic encryption scheme produces different ciphertexts when encrypting the same message multiple times. 
%As an example, the Goldwasser-Micali encryption scheme exploits quadratic residues to create pseudorandom ciphertexts for the same value.
The randomness brought by the probabilistic encryption scheme is essential to defend against the frequency analysis attack \cite{kerschbaum2015frequency}. 
In this paper, we design a FD-preserving, frequency-hiding encryption scheme based on probabilistic encryption. We use the private probabilistic encryption scheme based on pseudorandom functions \cite{katz2007introduction}. 
Specifically, for any plaintext value $p$, its ciphertext $e=<r, F_k(r)\oplus p>$, where $r$ is a random string of length $\lambda$, $F$ is a pseudorandom function, $k$ is the key, and $\oplus$ is the {\em XOR} operation. 

\vspace{-0.05in}
\subsection{Attacks and Security Model}
\label{sc:attack}
In this paper, we consider the {\em curious-but-honest} server, i,e., it follows the outsourcing protocols honestly (i.e., no cheating on storage and computational results), but it is curious to extract additional information from the received dataset. 
Since the server is potentially untrusted, the data owner sends the encrypted data to the server. 
The data owner considers the true identity of every cipher value as the sensitive information 
which should be protected. 
%We consider that the attacker may possess the following knowledge of $D$: (1) the domain values in $D$, and (2) the frequency distribution of data values in $D$. In reality, 

\noindent{\bf Frequency analysis attack.} We consider that the attacker may possess the frequency distribution knowledge of data values in $D$. In reality,  
the attacker may possess approximate knowledge of the value frequency in $D$. However, in order to make the analysis robust, we adopt the 
conservative assumption that the attacker knows the exact frequency 
of every plain value in $D$, and tries to break the encryption scheme by utilizing such frequency knowledge. 
We formally define the following security game $Exp_{\mathcal{A}, \Pi}^{freq}$ on the encryption scheme $\Pi=(KeyGen, Encrypt, Decrypt)$ : 
%Formally, consider an adversary $\mathcal{A}^{freq}$ that uses a security game $Exp_{\mathcal{A}, \Pi}^{freq}$ and the frequency distribution knowledge to break the encryption scheme $\Pi=(KeyGen, Encrypt, Decrypt)$. The experiment is defined as follows:
\vspace{-0.05in}
\begin{ditemize}
\item A random key $k$ is generated by running $KeyGen$. A set of plaintexts $\mathcal{P}$ is encrypted by running $Encrypt(\mathcal{P}, k)$. A set of ciphertext values $\mathcal{E} \leftarrow Encrypt(\mathcal{P}, k)$ is returned. 
\item A ciphertext value $e$ is randomly chosen from $\mathcal{E}$. Let $p = Decrypt(e, k)$. Let $freq_{\mathcal{P}}(p)$ and $freq_{\mathcal{E}}(e)$ be the frequency of $p$ and $e$ respectively. Let $freq(\mathcal{P})$ be the frequency distribution of $\mathcal{P}$. Then $e$, $freq_{\mathcal{E}}(e)$ and $freq(\mathcal{P})$ is given to the adversary $\mathcal{A}^{freq}$.
\item $\mathcal{A}^{freq}$ outputs a value $p'\in \mathcal{P}$.
\item The output of the experiment is defined to be $1$ if $p'=Decrypt(e, k)$, and $0$ otherwise. We write $Exp_{\mathcal{A}, \Pi}^{freq}=1$ if the output is $1$ and in this case we say that $\mathcal{A}^{freq}$ succeed. 
\end{ditemize}
\vspace{-0.05in}
\noindent{\bf Security Model.} To measure the robustness of the encryption scheme against the frequency analysis attack, we formally define {\em $\alpha$-security}. 
\begin{nameddefinition}{$\alpha$-security}
An encryption scheme $\Pi$ is $\alpha$-secure against frequency analysis attack if for every adversary $\mathcal{A}^{freq}$ it holds that 
\[Pr[Exp_{\mathcal{A}, \Pi}^{freq}=1]\leq \alpha,\]
where $\alpha$ is a value between $0$ and $1$.
\label{df:security}
\end{nameddefinition}

Intuitively, the smaller $\alpha$ is, the stronger security that $\Pi$ is against the frequency analysis attack. 
In this paper, we aim at designing an encryption scheme that provides $\alpha$-security for any given $\alpha\in (0,1)$.

\noindent{\bf Kerckhoffs's principle.} Kerckhoffs's principle \cite{shannon1949communication} requires that a cryptographic system should be secure even if everything about the system, except the key, is public knowledge.  Thus, we also require the encryption scheme to satisfy $\alpha$-security under Kerckhoffs's principle (i.e., the adversary knows the details of $F^2$ encryption algorithm). 

%Besides the frequency distribution, the attacker may know the details of the $F^2$ encryption scheme, and try to utilize such knowledge to break the cipher. Based on these two types of adversary knowledge, we consider two types of attacks, namely {\bf frequency analysis attack} ($Freq+\hat{D}$) and {\bf known-scheme attack} ($Freq+F^2$). More  details of these two attacks will be discussed in Section \ref{sc:security}.

\nop{
We assume that the attacker can be the compromised server. It may launch three types of attack, based on the various adversary knowledge it has:
\begin{itemize}
\item {\bf Freq+FD}: With the knowledge of the exact frequency information, together with the $FDs$ that are discovered from $D^k$, the attacker may try to re-construct the original dataset $D$. 

\item {\bf Freq+$D^k$}: The attacker can launch the {\em frequency  analysis attack} by matching the ciphertext values in $D^k$ with the plaintext values by their frequency. 

\item {\bf Freq+$F^2$}: The attacker knows the details of the encryption scheme, and tries to break the cipher by utilizing the knowledge of the encryption scheme. 
\end{itemize}
}

\nop{
\vspace{-.05in}
\subsection{Security Model}
\label{sc:model}
In order to measure the success probability of the frequency analysis attack to break our encryption scheme, we design an experiment $Exp_{\mathcal{A}, \Pi}^{freq}$ that considers an adversary $\mathcal{A}^{freq}$ with frequency distribution knowledge and an encryption scheme $\Pi=(KeyGen, Encrypt, Decrypt)$. The experiment is defined as follows:

\begin{ditemize}
	\item 1. A random key $k$ is generated by running $KeyGen$. A set of plaintexts $\mathcal{P}$ is encrypted by running $Encrypt(\mathcal{P}, k)$. A set of ciphertext values $\mathcal{E} \leftarrow Encrypt(\mathcal{P}, k)$ is returned. 
	\item 2. A ciphertext value $e$ is randomly chosen from $\mathcal{E}$. 
	\item 3. Let $freq_{\mathcal{P}}(p)$ and $freq_{\mathcal{E}}(e)$ be the frequency of $p$ in $\mathcal{P}$ and $e$ in $\mathcal{E}$ respectively. Let $freq(\mathcal{P})$ be the frequency distribution of $\mathcal{P}$. Then $e$, $freq_{\mathcal{E}}(e)$ and $freq(\mathcal{P})$ is given to the adversary $\mathcal{A}^{freq}$.
	\item 4. $\mathcal{A}^{freq}$ outputs a value $p'\in \mathcal{P}$.
	\item 5. The output of the experiment is defined to be $1$ if $p'=Decrypt(e, k)$, and $0$ otherwise. We write $Exp_{\mathcal{A}, \Pi}^{freq}=1$ if the output is $1$ and in this case we say that $\mathcal{A}^{freq}$ succeed. 
\end{ditemize}

\begin{nameddefinition}{$\alpha$-security}
An encryption scheme $\Pi$ is $\alpha$-secure against frequency analysis attack if for every adversary $\mathcal{A}^{freq}$ it holds that 
\[Pr[Exp_{\mathcal{A}, \Pi}^{freq}=1]\leq \alpha,\]
where $\alpha$ is a value between $0$ and $1$.
\label{df:security}
\end{nameddefinition}

In this paper, we aim at designing an encryption scheme that provides $\alpha$-security for any given $\alpha\in (0,1)$.

So far, we only consider the security model based on the frequency distribution knowledge. Besides that, the Kerckhoffs's principle \cite{shannon1949communication} demands that the security of an encryption scheme should solely depend on the secrecy of the key, but not the obscurity of the system. In other words, it forces security against the known-scheme attack, in which the adversary can know every detail about our encryption scheme. Thus, we also require the encryption scheme to satisfy $\alpha$-security under Kerckhoffs's principle. 
}
\nop{
We consider the attack that tries to construct the mapping between the plaintext and ciphertext values. In particular, given a set of plaintext values $\cal P$ and a set of ciphertext values $\cal E$, for each cipher value $e\in\cal E$, the attacker constructs a set of candidate plaintext values $Cand(e) \subseteq \cal P$. The probability of mapping a cipher value $e$ to its plaintext value $p$ is 
$prob(e\rightarrow p) = 1 /|Cand(e)|$. To quantify our security guarantee, we define {\em $\alpha$-security}. 
\vspace{-.05in}
\begin{nameddefinition}{$\alpha$-security}
{\em
given a set of plaintext values $\cal P$ and a set of ciphertext values $\cal E$, we say $\cal E$ satisfies {\em $\alpha$-security} if for each cipher value $e\in\cal E$, $prob(e\rightarrow p) \leq \alpha$, where $\alpha$ is a user specified threshold. 
}
\end{nameddefinition}
\vspace{-.05in}
We will analyze the security guarantee of our $F^2$ encryption scheme against both {\bf Freq+$\hat{D}$} and {\bf Freq+$F^2$} attacks in Section \ref{sc:security}.
}

\vspace{-0.05in}
\section{FD-preserving Encryption}
\label{sc:encryption}

%Intuitively, any simple one-to-one substitution scheme that is applied at the cell level is always $FD$-preserving. However, such encryption scheme is vulnerable against the frequency analysis attack, as the attacker can easily map the cipher values to plain ones based on their frequency.  

\nop{
It has been shown in Introduction (Section 1) that one-to-one substitution schemes are vulnerable against the frequency analysis attack.  An effective solution to defend against the frequency analysis attack is to apply the {\emm homophonic substitution} (probabilistic encryption) scheme, by which the multiple occurrences of the same plaintext value are replaced with different ciphertext symbols, so that the frequency distribution of ciphertext values does not match that of the plaintext values. 
Now the challenge is to find on which attributes to apply the probabilistic encryption. A straightforward solution is to apply the probabilistic encryption scheme on each attribute individually. Specifically, for each attribute $A$, the probabilistic encryption scheme is applied on its values $a_1, \dots, a_t$, such that multiple occurrences of $a_i\ (1\leq i\leq t)$ are encrypted to be different cipher values.
}

The key to design a FD-preserving encryption algorithm is to first identify the set of attributes on which the probabilistic encryption scheme is applied. We have the following theorem to show that for any FD $F$, if the probabilistic encryption scheme is applied on the attribute set that includes both $LHS(F)$ and $RHS(F)$, $F$ is still preserved in the encrypted data. 
\vspace{-.05in}
\begin{theorem}
\label{theorem:1}
Given a dataset $D$ and any FD $F$ of $D$, let the attribute set $\mathcal{A}$ be the attribute sets on which the probabilistic encryption scheme  is applied on, and let $\hat{D}$ be the encryption result. Then $F$ always holds in $\hat{D}$ if $LHS(F)\cup RHS(F)\subseteq \mathcal{A}$. 
\end{theorem}
\begin{proof}
Assume that there is a FD: $A \rightarrow B$ which holds in $D$ but not in $\hat{D}$. It must be true that there exists at least one pair of records $r_1$, $r_2$ such that $r_1[A]=r_2[A]$ and $r_1[B]=r_2[B]$ in $D$, while in $\hat{D}$, $\hat{r_1}[A]=\hat{r_2}[A]$ but $\hat{r_1}[B]\neq \hat{r_2}[B]$. According to the HS scheme which is applied on an attribute set $\cal{A}$ s.t. $LHS(F)\cup RHS(F) \subseteq \cal{A}$, there are two cases. First, if $r_1$ and $r_2$ are taken as the same instance, then $\hat{r_1}$ and $\hat{r_2}$ have the same value in every attribute. It cannot happen as $\hat{r_1}[B]\neq \hat{r_2}[B]$. Second, if $r_1$ and $r_2$ are taken as different instances, then it must be true that $\hat{r_1}[A] \neq \hat{r_2}[A]$ and $\hat{r_1}[B] \neq \hat{r_2}[B]$ according to Requirement 2 of the HS scheme. So $\hat{r_1}$ and $\hat{r_2}$ cannot break the FD: $A \rightarrow B$ in $\hat{D}$. 
\end{proof}

\vspace{-.05in}
To continue our running example, consider the base table in Figure \ref{table:release} (a). Figure \ref{table:release} (d) shows an example table $\hat{D}_3$ of applying the probabilistic encryption scheme on the attribute set $\{A, B\}$. By this scheme, the four instances of $(a_1, b_1)$ are encrypted as $(\hat{a}_1^1, \hat{b}_1^1$) and $(\hat{a}_1^2, \hat{b}_1^2$). Now the FD $A\rightarrow B$ still holds on $\hat{D}_3$. 

Based on Theorem \ref{theorem:1}, we formally define our FD-preserving probabilistic encryption scheme. We use $|\sigma_{\mathcal{A}=r[\mathcal{A}]}(D)|$ to specify the number of records in $D$ that have the same value as $r[\mathcal{A}]$, for a specific record $r$ and a set of attributes $\mathcal{A}$. 
\vspace{-.05in}
\begin{nameddefinition}{FD-preserving probabilistic encryption scheme}
\label{def:hs}
Given a table $D$, a FD $F$ of $D$, and a set of attributes $\mathcal{A} = \{A_1, \dots, A_g\}$ of $D$ such that $LHS(F)\cup RHS(F) \subseteq \mathcal{A}$, for any instance $r \{a_1, \dots, a_g\}\in D$, let 
$f=|\sigma_{\mathcal{A}=r[\mathcal{A}]}(D)|$. When $f>1$, the FD-preserving probabilistic encryption scheme encrypts the $f$ instances as $t>1$ unique instances $\{r_1, \dots, r_t\}$, such that $r_1 = \{\hat{a}_1^1, \dots, \hat{a}_g^1\}$ of frequency $f_1$, $\dots$, and $r_t = \{\hat{a}_1^t, \dots, \hat{a}_g^t\}$ of frequency $f_t$. We require that: 
\vspace{-0.05in}
\begin{ditemize}
\item Requirement 1: $\sum_{i=1}^{t} f_i = f$; 
\item Requirement 2: $\forall i\ (1\leq i\leq g)$, $\hat{a}_i^x\neq\hat{a}_i^y$, for all $x, y\in[1, t]$ where $x \neq y$. 
\end{ditemize}
\vspace{-0.05in}
%It is not necessary that $f_i$ and $f_j$ are the same, for all $i, j\in [1, t]$. 
\end{nameddefinition}
\vspace{-.05in}
Requirement 1 requires when encrypting a plaintext value set to multiple different ciphertext instances, the sum of frequency of these ciphertext instances is the same as the frequency of the plaintext value set. Requirement 2 requires that the ciphertext values of the same plaintext value set do not overlap at any single attribute. The aim of Requirement 2 is to provide $\alpha$-security under Kerckhoffs's principle (more details in Section \ref{sc:mali}). We require that $f>1$, since the probabilistic encryption scheme becomes a deterministic encryption scheme when $f=1$.
%, and thus not safe against the frequency analysis attack. 

Our FD-preserving probabilistic encryption method consists of four steps: (1) finding maximum attribute sets; (2) splitting-and-scaling; (3) conflict resolution; and (4) eliminating false positive FDs. We explain the details of these four steps in the following subsections. 

\vspace{-0.05in}
\subsection{Step 1: Finding Maximum Attribute Sets}

Theorem \ref{theorem:1} states that the set of attributes on which the probabilistic encryption scheme is applied should contain all attributes in FDs. Apparently the set of all attributes of $D$ satisfies the requirement. However, it is not a good solution since it is highly likely that there does not exist a probabilistic encryption scheme, due to the reason that $f=|\sigma_{\mathcal{A}=r[\mathcal{A}]}(D)|$ is more likely to be 1 when $\mathcal{A}$ contains more attributes. Now the main challenge is to decide the appropriate attribute set $\mathcal{A}$ that the probabilistic encryption scheme will be applied on, assuming that the data owner is not aware of the existence of any $FD$. To address this challenge, we define the {\em maximum attribute set} on which there exists at least one instance whose frequency is greater than 1. Formally, 
\vspace{-.05in}
\begin{nameddefinition}{Maximum Attribute Set ($MAS$)}
Given a dataset $D$, an attribute set $\mathcal{A}$ of $D$ is a {\em maximum attribute set} $MAS$ if: 
(1) there exists at least an instance $\mathcal{a}$ of $\mathcal{A}$ such that $|\sigma_{\mathcal{A}=\mathcal{a}}(D)| > 1$; and 
(2) for any attribute set $A'$ of $D$ such that $\mathcal{A}\subseteq \mathcal{A}'$, there does not exist an instance $\mathcal{a}'$ of $\mathcal{A}'$ s.t. $|\sigma_{\mathcal{A}'=\mathcal{a}'}(D)|> 1$. 
\end{nameddefinition}
\vspace{-.05in}
Our goal is to design the algorithm that finds {\em all} $MASs$ of a given dataset $D$. Note that the problem of finding $MASs$ is not equivalent to finding FDs. For instance, consider the base table $D$ in Figure \ref{table:release} (a). Its $MASs$ is \{A, B, C\}. But its FD is $A\rightarrow B$. In general, given a set of $MAS$s $\cal M$ and a set of FDs $\cal F$, for each FD $F\in\cal F$, there always exists at least an $MAS$ $M\in\cal M$ such that $(LHS(F)\cup RHS(F))\subseteq M$. 
%We will show that the complexity of finding $MASs$, which is $O(nm)$, is cheaper than the complexity of finding FDs ($O((n+m^{2.5})2^m$)  \cite{huhtala1999tane}, where $m$ is the number of attributes, and $n$ is the number of tuples. 

In general, finding $MASs$ is quite challenging given the exponential number of attribute combinations to check. We found out that our $MAS$ is equivalent to the {\em maximal non-unique column combination} \cite{heise2013scalable}. Informally, maximal non-unique column combination refers to a set of columns whose projection has duplicates. It has been shown that finding all (non-)unique column combination is an NP-hard problem \cite{gunopulos2003discovering}. In \cite{heise2013scalable}, the authors designed a novel algorithm named $Ducc$ that can discover maximal non-unique column combinations efficiently. The complexity of $Ducc$ is decided by the solution set size but not the number of attributes. 
Therefore, we adapt the $Ducc$ algorithm \cite{heise2013scalable} to find $MASs$. Due to the space limit, we omit the details of the algorithm here. For each discovered $MAS$, we find its {\em partitions} \cite{huhtala1999tane}. We say two tuples $r$ and $r'$ are {\em equivalent} with respect to a set of attributes $X$ if $r[X] =r'[X]$. 
\vspace{-.05in}
\begin{nameddefinition}{Equivalence Class ($EC$) and Partitions}\cite{huhtala1999tane}
The {\em equivalence class} ($EC$) of a tuple $r$ with respect to an attribute set $X$, denoted as $r_X$, is defined as $r_X = \{r'| r[A] =r'[A], \forall A\in X\}$. 
The {\em size} of $r_X$ is defined as the number of tuples in $r_X$,  and the {\em representative value} of $r_X$ is defined as $r[X]$. 
The set $\pi_X = \{r_X|r\in D\}$ is defined as a {\em partition} of $D$ under the attribute set $X$. That is, $\pi_X$ is a collection of disjoint sets ($ECs$) of tuples, such that each set has a unique representative value of  a set of attributes $X$, and the union of the sets equals $D$. 
\end{nameddefinition}
\vspace{-.05in}
%Note that $r_X$ includes $r$ itself. 
As an example, consider the dataset $\hat{D}_3$ in Figure \ref{table:release} (d), $\pi_{\{A, B\}}$ consists of two $EC$s whose  representative values are 
$\{\hat{a}_{1}^{1}, \hat{b}_{1}^{1}\}$ and $\{\hat{a}_{1}^{2}, \hat{b}_{1}^{2}\}$. Apparently, a $MAS$ is an attribute set whose partitions contain at least one equivalence class whose size is more than 1. 

%%%%%%%%%%%%%%%%%%%old version%%%%%%%%%%%%%%%%%%%%%%%%
%%%%%%%%%%%%%%%%%%%old version%%%%%%%%%%%%%%%%%%%%%%%%
\nop{
\vspace{-0.05in}
\subsection{Step 1: Finding Maximum Attribute Sets}

The main challenge is to decide on which attributes the probabilistic encryption scheme will be applied on, assuming that the data owner is not aware of the existence of any $FD$. To address this challenge, we define the {\em maximum attribute set} on which there exists at least one instance whose frequency is greater than 1. We use $D[A]$ to denote the projection on attribute(s) $A$, $D[A=a]$ to denote the set of instances of $D$ whose attribute $A$ is valued $a$, and $f(D[A=a])$ to denote the number of such instances. Formally, 
given a dataset $D$, an attribute set $A$ of $D$ is a {\em maximum attribute set} $MAS$ if: 
(1) there exists at least an instance $a$ of $A$ s.t. $f(D[A=a])\geq 1$, and 
(2) for all attribute set $A'$ s.t. $A\subseteq A'$, there does not exist an instance $a'\in D$ s.t. $f(D[A'=a'])\geq 1$. 

Our goal is to design the algorithm to find all maximum attribute sets of a given dataset $D$. Note that finding the maximum attribute sets is not equivalent to finding FDs. In general, given a set of $MAS$s $\cal M$ and a set of FDs $\cal F$, for each FD $F\in\cal F$, there always exists at least an $MAS$ $M\in\cal M$ such that $(LHS(F)\cup RHS(F))\subseteq M$. 
The complexity of finding $MASs$, which is $O(nm)$, is cheaper than the complexity of finding FDs ($O((n+m^{2.5})2^m$)  \cite{huhtala1999tane}, where $m$ is the number of attributes, and $n$ is the number of tuples. 

We use the concept of {\em partitions} \cite{huhtala1998efficient} to find the $MASs$. Formally, we say two tuples $r$ and $r'$ are {\em equivalent} with respect to a set of attributes $X$ if $r[X] =r'[X]$.  We define the {\em equivalence class} ($EC$) of a tuple $r$ with respect to an attribute set $X$, denoted as $r_X$, 
% by $[T]_X$, 
as $r_X = \{r'| r[A] =r'[A], \forall A\in X\}$. 
%the set of tuples that are equivalent to $r$ with respect to $X$ . 
%, i.e., $[T]_X = \{T'| T[A] =T'[A], \forall A\in X\}$. 
Note that $r_X$ includes $r$ itself. We define the {\em size} of $r_X$ as the number of tuples in $r_X$, and $r[X]$ as the {\em representative value} of $r_X$. We define the set $\pi_X$ of $EC$s as a {\em partition} of $D$ under the attribute set $X$, in which each $EC$ corresponds to a unique representative value of $X$. 
For example, consider the dataset $D$ in Figure \ref{table:release} (a), $\pi_{\{A, B\}}$ consists of one $EC$ whose  representative value is $\{a_1, b_1\}$. Apparently, a $MAS$ is an attribute set whose partitions contain at least one equivalence class whose size is more than 1. 
%, if (1) for any two equivalence classes of $\pi_X$, their representative values do not overlap on any attribute of $X$; and (2) the union of all equivalence classes of $\pi_X$ covers $D$. 

\begin{figure}[t!]
\begin{center}
\vspace{-0.1in}
\includegraphics[width=0.3\textwidth]{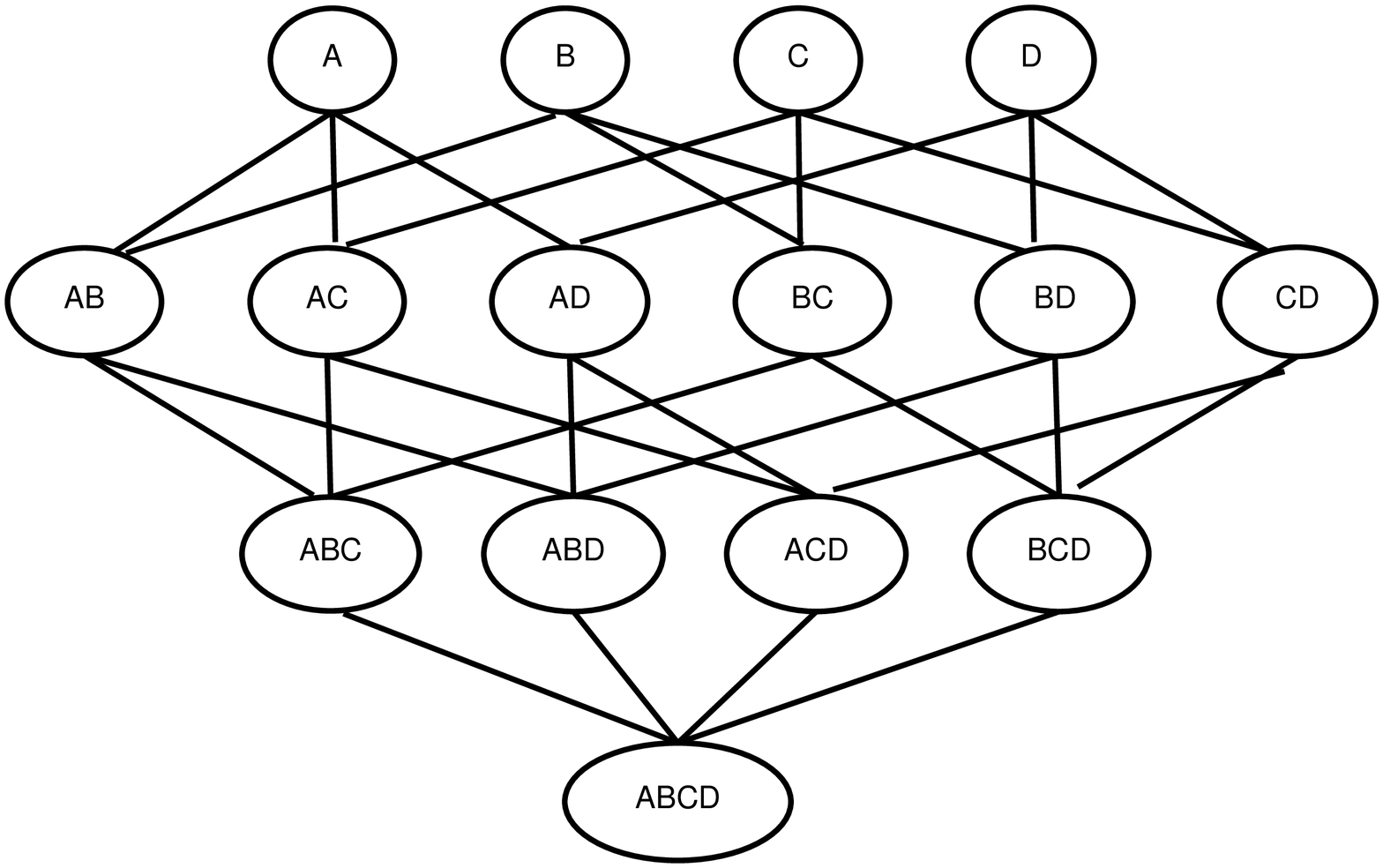}
\vspace{-0.05in}
\caption{\label{fig:lattice}An example of attribute lattice}
\end{center}
\vspace{-0.2in}
\end{figure}

Our algorithm of finding the $MASs$ relies on a data structure called {\em attribute lattice}. The lattice is constructed in a top-down fashion; each node at the top of the lattice corresponds to a single attribute of $D$. The bottom of the lattice consists of only one node that corresponds to all attributes of $D$. Each node at the $i^{th} (i>1)$ level corresponds to an attribute set that consists of $i$ attributes. There is an edge between an $i^{th}$-level node $N$ and an $(i+1)^{th}$-level node $N'$ if the attribute set of $N$ is a subset of the attribute set of $N'$. We call node $N$ a {\em parent} of $N'$, and $N'$ a {\em child} of $N$. Figure \ref{fig:lattice} shows an example of the attribute lattice. Node $\{A, B\}$ is a parent of the node $\{A, B, C\}$. It is possible that a node (except the node at the bottom) has multiple children, and a node (except the node at the top) has multiple parents. 

Our algorithm starts from the top of the attribute lattice. We pick a node $N$ from the top of the lattice in a random fashion, and compute the partitions with regard to the attribute(s) that the picked lattice node corresponds to. Let the picked attribute be $A$. We initialize $X=\{A\}$, and compute $\pi_{X}$. If all equivalence classes of $\pi_{X}$ are of size one, we mark the corresponding lattice node of $X$ as ``non-candidate'' and move to the next sibling of $N$ in the lattice. Otherwise, we mark the corresponding lattice node $N$ of $X$ as ``candidate'', mark all parent nodes of $N$ as ``candidate'' as well, and move to a child node of $N$ in the lattice. We follow the random fashion to pick the child node if there are multiple children. For each node that is marked as ``candidate'' but all of its children are ``non-candidate'', we return it as a $MAS$. We repeat this procedure until all nodes in the lattice are marked. One advantage of the partitions is that $\pi_{X\cup Y}$ can be computed from $\pi_{X}$ and $\pi_{Y}$ efficiently \cite{huhtala1998efficient}. Therefore, we only need to scan the dataset $m$ times to calculate $\pi$ for $m$ attributes individually. The complexity of finding all $MASs$ is $O(nm)$. 
}

\vspace{-0.05in}
\subsection{Step 2: Splitting-and-scaling Encryption}
\label{sc:sd}
After the $MASs$ and their partitions are discovered, we design two steps to apply the probabilistic encryption scheme on  the partitions: (1) grouping of $ECs$, and (2) splitting and scaling on the $EC$ groups. Next, we explain the details of the two steps.  
%Intuitively, given a $MAS$ $X$ and its partition $\pi_{X}$, we randomly pick $\ell= [\frac{1}{\alpha}]$ equivalence classes of $\pi_{X}$, where $\alpha$ is the threshold for $\alpha$-security. Let the representative values of the picked equivalence classes be $C_1, \dots, C_{\ell}$ and their frequency be $f_1, \dots, f_{\ell}$.  We apply the following two steps to apply the probabilistic encryption scheme on these equivalence classes.

\vspace{-0.05in}
\subsubsection{Step 2.1. Grouping of Equivalence Classes} 

To provide $\alpha$-security, we group the $EC$s in the way that each equivalence class belongs to one single group, and each group contains at least $k\geq [\frac{1}{\alpha}]$ $EC$s, where $\alpha$ is the threshold for $\alpha$-security. We use $ECG$ for the {\em equivalence class group} in short. 

We have two requirements for the construction of $ECGs$. First, we prefer to put $EC$s of close sizes into the same group, so that we can minimize the number of new tuples added by the next splitting \& scaling step (Section \ref{sc:s&s}). Second, we do not allow any two $EC$s in the same $ECG$ to have the same value on any attribute of $MAS$. Formally, 
\vspace{-.05in}
\begin{nameddefinition}{Collision of $ECs$}
Given a $MAS$ $M$ and two equivalence classes $C_i$ and $C_j$ of $M$, we say $C_i$ and $C_j$ have {\em collision} if there exists at least one attribute $A\in M$ such that $C_i[A]=C_j[A]$. 
\end{nameddefinition}
\vspace{-.05in}
For security reason, we require that all $ECs$ in the same $ECG$ should be collision-free (more details in Section \ref{sc:semi}). 

To construct $ECGs$ that satisfy the aforementioned two requirements, first, we sort the equivalence classes by their sizes in ascending order. Second, we group the collision-free $ECs$ with the closest size into the same $ECG$, until the number of non-collisional $ECs$ in the $ECG$ reaches $k =  [\frac{1}{\alpha}]$. It is possible that for some $ECG$s, the number of $ECs$ that can be found is less than the required $k = [\frac{1}{\alpha}]$. In this case, we add {\em fake} collision-free $ECs$ to achieve the size $k$ requirement. The fake $ECs$ only consist of the values that do not exist in the original dataset. The size of these fake $ECs$ is set as the minimum size of the $ECs$ in the same $ECG$. We must note that the  server cannot distinguish the fake values from real ones, even though it may be aware of some prior knowledge of the outsourced dataset. This is because both true and fake values are encrypted before outsourcing.  
 %The complexity of $ECG$ construction is $O(plgp)$, where $p$ is the number of equivalence classes of $D$. 

\nop{
Algorithm \ref{alg:grouping} specifies the detailed procedure to generate $ECGs$.

\begin{algorithm}
\begin{small}
\caption{ $Grouping(\cal C$, $k$)
\label{alg:grouping}
}
\begin{algorithmic}[1]
\REQUIRE{the set of  equivalence groups $\cal C$ $=\{C_1, \dots, C_{\ell}\}$ in partition $\pi_X$, the minimal $ECG$ size $k$}
\ENSURE{Every equivalence class is grouped into a valid $ECG$.}
\STATE{Sort $C_i(1\leq i \leq l)$ in descending order by the equivalence class's frequency;}
\FOR{$i=1$ \TO $l$} 
  \IF{$C_i$ is not grouped} 
    \STATE{$G$ $=$ $new$ $ECG()$;}
    \FORALL{$A_p \in X$} 
      \STATE{$G.A_p = \{C_i.A_p\}$;}
    \ENDFOR
    \FOR{$j=i+1$ \TO $l$}
      \IF{($C_j$ is not grouped)} 
        \STATE{$flag=TRUE$;}
         \FORALL{$A_p \in X$} 
          \IF{$C_j.A_p \in G.A_p$}
            \STATE{$flag=FALSE$;}
            \BREAK
          \ENDIF
         \ENDFOR
         \IF{$flag==TRUE$}
          \STATE{$G=G\cup \{C_j\}$;}
         \ENDIF
         \IF{$G.size \geq k$}
          \BREAKBesides, normally $|M|$ is much smaller than $m$, where $m$ is the number of attributes in $D$.

         \ENDIF
      \ENDIF
    \ENDFOR
  \ENDIF
\ENDFOR
\FORALL{$G \in \{ECG|ECG.size < k\}$}
  \STATE{Add {\em fake} collision-free equivalence classes to make $G.size \geq k$;}
\ENDFOR
\end{algorithmic}
\end{small}
\end{algorithm} 
}

\vspace{-0.05in}
\begin{figure}[ht]
\begin{center}
\begin{small}
\begin{tabular}{| c | c | c | c|}
  \hline
  ID & Representative value & Equivalence class & Size\\ \hline
  $C_1$ & $(a_1, b_1)$ & $\{r_1, r_4, r_5, r_7, r_{12}\}$ & 5\\ \hline
	$C_2$ & $(a_1, b_2)$ & $\{r_2, r_6, r_8, r_{14}\}$ & 4\\ \hline
  $C_3$ & $(a_2, b_2)$ & $\{r_3, r_9, r_{16}\}$ & 3\\ \hline
  $C_4$ & $(a_2, b_1)$ & $\{r_{10}, r_{11}\}$ & 2\\ \hline
  $C_5$ & $(a_3, b_3)$ & $\{r_{13}, r_{15}\}$ & 2\\ \hline  
\end{tabular}
\vspace{-.1in}
\caption{\label{fig:ecg}An example of $ECG$ construction}
\end{small}
\end{center}
\vspace{-.3in}
\end{figure}

As an example, consider an $MAS$ $M=\{A, B\}$ and its five $ECs$ shown in Figure \ref{fig:ecg}. Assume it is required to meet $\frac{1}{3}$-security (i.e., $\alpha = \frac{1}{3}$). The two $ECG$s are $ECG_1=\{C_1, C_3, C_6\}$, and $ECG_2=\{C_2, C_4, C_5\}$, where $C_6$ (not shown in Figure \ref{fig:ecg}) is a fake $EC$ whose representative value is $(a_4, b_4)$ ($a_4$ and $b_4$ do not exist in the original dataset). Both $ECG_1$ and $ECG_2$ only have collision-free $ECs$, and each $ECG$ contains at least three $ECs$. Note that $C_1$ and $C_2$ cannot be put into the same $ECG$ as they share the same value $a_1$, similarly for $C_2$ and $C_3$ as well as for $C_3$ and $C_4$. 

\nop{
  \begin{figure*}[ht]
\begin{center}
\begin{small}
\begin{tabular}{c c c c c}
\begin{tabular}{| c | c | c | c |}
\hline
  & A & B & C \\ \hline
  $r_1$ & $a_1$ & $b_1$ & $c_1$ \\
  $r_2$ & $a_3$ & $b_1$ & $c_1$ \\
  $r_3$ & $a_1$ & $b_2$ & $c_2$ \\
  $r_4$ & $a_2$ & $b_2$ & $c_2$ \\
  $r_5$ & $a_2$ & $b_2$ & $c_3$ \\
  $r_6$ & $a_2$ & $b_2$ & $c_4$ \\
  $r_7$ & $a_2$ & $b_2$ & $c_5$ \\ \hline  
\end{tabular}
& 
\begin{tabular}{| c | c | c | c |}
\hline
   & A & B & C \\ \hline
  $r_1$ & & & \\
  $r_2$ & & & \\
  $r_3$ & & & \\
  $r_4$ & $\hat{a}_2^1$ & $\hat{b}_2^1$ & \\
  $r_5$ & $\hat{a}_2^1$ & $\hat{b}_2^1$ & \\
  $r_6$ & $\hat{a}_2^1$ & $\hat{b}_2^1$ & \\
  $r_7$ & $\hat{a}_2^1$ & $\hat{b}_2^1$ & \\ \hline
\end{tabular}
&
\begin{tabular}{| c | c | c | c |}
\hline
   & A & B & C  \\ \hline
  $r_1$ & & $\hat{b}_1^1$ & $\hat{c}_1^1$ \\  
  $r_2$ & & $\hat{b}_1^1$ & $\hat{c}_1^1$ \\  
  $r_3$ & & $\hat{b}_2^2$ & $\hat{c}_2^1$ \\  
  $r_4$ & & $\hat{b}_2^2$ & $\hat{c}_2^1$ \\  
  $r_5$ & & & \\
  $r_6$ & & & \\
  $r_7$ & & & \\  
  \hline
\end{tabular}
&
\begin{tabular}{| c | c | c | c |}
  \hline
   & A & B & C \\ \hline
  $r_1$ &  & $\hat{b}_1^1$ & $\hat{c}_1^1$ \\
  $r_2$ &  & $\hat{b}_1^1$ & $\hat{c}_1^1$ \\
  $r_3$ &  & $\hat{b}_2^2$ & $\hat{c}_2^1$ \\
  $r_4$ & $\hat{a}_2^1$ & $\hat{b}_2^1$ / $\hat{b}_2^2$ & $\hat{c}_2^1$ \\
  $r_5$ & $\hat{a}_2^1$ & $\hat{b}_2^1$ &  \\
  $r_6$ & $\hat{a}_2^1$ & $\hat{b}_2^1$ &  \\
  $r_7$ & $\hat{a}_2^1$ & $\hat{b}_2^1$ &  \\ \hline
\end{tabular}
&
\begin{tabular}{| c | c | c | c |}
  \hline
   & A & B & C \\ \hline
  $r_1$ & $\hat{a}_1^1$ & $\hat{b}_1^1$ & $\hat{c}_1^1$ \\
  $r_2$ & $\hat{a}_3^1$ & $\hat{b}_1^1$ & $\hat{c}_1^1$ \\
  $r_3$ & $\hat{a}_1^1$ & $\hat{b}_2^1$ & $\hat{c}_2^1$ \\
  $r_4$ & $\hat{a}_2^1$ & $\hat{b}_2^1$ & $\hat{c}_2^1$ \\
  $r_5$ & $\hat{a}_2^1$ & $\hat{b}_2^1$ & $\hat{c}_3^1$ \\
  $r_6$ & $\hat{a}_2^1$ & $\hat{b}_2^1$ & $\gamma_4^1$ \\
  $r_7$ & $\hat{a}_2^1$ & $\hat{b}_2^1$ & $\gamma_5^1$ \\ \hline
\end{tabular}
\\ 
(a) Original dataset 
&
(b) $Enc_X(D)$
&
(c) $Enc_Y(D)$
&
(d) Conflictive encryption
&
(e) Syrhonized encryption
\\
\end{tabular}
\caption{\label{fig:sync1} An example of synchronization}
\end{small}
\end{center}
\end{figure*}
}

\vspace{-0.05in}
\subsubsection{Step 2.2. Splitting-and-scaling ($S \& S$)} 
\label{sc:s&s}
This step consists of two phases, {\em splitting} and {\em scaling}. In particular, consider an $ECG$ $\overline{\cal C}$ = $\{C_1, \dots, C_x\}$, in which each $EC$ $C_i$ is of size $f_i$ $(1\leq i \leq x)$. By {\em splitting}, for each $C_i$, its $f_i$ (identical) plaintext values are encrypted to ${\varpi}$ unique ciphertext values, each of frequency $[\frac{f_i}{\varpi}]$, where $\varpi$ is the split factor whose value is specified by the user. The {\em scaling} phase is applied after splitting. By scaling, all the ciphertext values reach the same frequency by adding additional copies up to $[\frac{f_{max}}{\varpi}]$, where $f_{max}$ is the maximum size of all $ECs$ in $\overline{\cal C}$. After applying the splitting \& scaling ($S \& S$), all ciphertext values in the same $ECG$ are of the same frequency. 

It is not necessary that every $EC$ must be split. Our aim is to find a subset of given $ECs$ whose split will result in the minimal amounts of additional copies added by the scaling phase. In particular, given an $ECG$ $\overline{\cal C}$ = $\{C_1, \dots, C_k\}$ in which $ECs$ are sorted  by their sizes in ascending order (i.e., $C_k$ has the largest size), we aim to find the {\em split point} $j$ of $\overline{\cal C}$ such that each $EC$ in \{$C_1, \dots, C_{j-1}\}$ is not split but each $EC$ in \{$C_j, \dots, C_k\}$ is split. We call $j$ the {\em split point}. 
% the number of duplicated records we need to add is $R=\sum\limits_{i=1}^{k}(f_k-f_i)$.  
Next, we discuss how to find the {\em optimal} split point that delivers the minimal amounts of additional copies by the scaling phase. There are two cases: (1) the size of $C_k$ is still the largest after the split; and (2) the size of $C_k$ is 
not the largest anymore, while the size of $C_{j-1}$ (i.e., the EC of the largest size among all ECs that are not split) becomes the largest. We discuss these two cases below. 

\indent{\bf Case 1:} $[\frac{f_k}{\varpi}] \geq f_{j-1}$, i.e., the split of $C_k$ still has the largest frequency within the group. In this case, the total number of copies added by the scaling step is 
%$R_1= \sum\limits_{i=1}^{j-1} ([\frac{f_k}{\varpi}] - f_i) + {\varpi}\sum\limits_{i=j}^{k}([\frac{f_k}{\varpi}] - [\frac{f_i}{\varpi}]) = \sum\limits_{i=1}^{j-1} (\varpi f_k - f_i) + \sum\limits_{i=j}^{k}(f_k - f_i)$. 
\vspace{-.05in}
\[R_1 = \sum\limits_{i=1}^{j-1} ([\frac{f_k}{\varpi}] - f_i) + \sum\limits_{i=j}^{k}(f_k - f_i).\]
%The reduced number of duplicates is $\Delta_1=R-R_1=\sum\limits_{i=1}^{j-1}(f_k-[\frac{1}{\varpi}*f_k])=(j-1)(1-\frac{1}{\varpi})*f_k$. It's obvious that to maximize $\Delta_1$ in order to minimize the number of added records. 
It can be easily inferred that when $j=max\{ j|f_{j-1}\leq [\frac{f_k}{\varpi}]\}$, $R_1$ is minimized. 
%The corresponding $\Delta_1^{max}=(j_{max}-1)*(1-\frac{1}{\varpi})*f_k$ is the optimized result under this case.

\indent{\bf Case 2:} $[\frac{f_k}{\varpi}] < f_{j-1}$, which means that $C_{j-1}$ enjoys the largest frequency after splitting. For this case, the number of duplicates that is to be added is: 
\vspace{-.05in}\[R_2=\sum\limits_{i=1}^{j-1}(f_{j-1}-f_i)+\varpi\sum\limits_{i=j}^k(f_{j-1}-[\frac{f_i}{\varpi}]).\] 
%The reduced number of duplicates is $\Delta_2=R-R_2=k*f_k-({\varpi}*k-({\varpi}-1)*(j-1))*f_{j-1}$. 
 %When $j=max\{ j|f_{j-1}\leq \varpi f_k \}$, the value $R_2$ will be minimized. 
$R_2$ is not a linear function of $j$. Thus we define $j_{max} = max\{j|[\frac{f_k}{\varpi}] > f_{j-1}\}$. We try all $j\in[j_{max}, k]$ and return $j$ that delivers the minimal $R_2$. 

For any given $ECG$, the complexity of finding its optimal split point is $O(|ECG|)$. In practice, the optimal split point $j$ is close to the $ECs$ of the largest frequency (i.e., few split is needed). 

\nop{
Based on the analysis of the two cases, we design an algorithm to find the best split point in each equivalence class. The pseudo code is shown in Algorithm \ref{alg:findj}. 
\begin{algorithm}
  \caption{$FindSplitPoint(\overline{\cal C}$, $\varpi)$}
  \label{alg:findj}
  \begin{algorithmic}
    \REQUIRE{$\overline{\cal C}$ $=\{C_1, \dots, C_k\}$ and their frequency $f_1, \dots, f_k$, $\varpi$}
    \ENSURE {The best split point $j$ that incur minimal number of inserted duplicates}
    \STATE $j_{max} = max\{j|f_{j-1} \leq [\frac{f_k}{\varpi}]\}$;
    \STATE $\Delta_{j_{max}} = R-R_1=\sum\limits_{i=1}^{j_{max}-1}(f_k-\frac{f_k}{\varpi})=(j_{max}-1)(1-\frac{f_k}{\varpi})$;
    \FOR {$j = j_{max}+1 \to k$}
      \STATE $\Delta_j = R-R_2 = \sum\limits_{i=1}^{k}(f_k-f_i)-(\sum\limits_{i=1}^{j-1}(f_{j-1}-f_i)+{\varpi}\sum\limits_{i=j}^k(f_{j-1}-\frac{1}{\varpi} f_i)) = kf_k-({\varpi}k-({\varpi}-1)(j-1))f_{j-1}$;
    \ENDFOR
    \STATE $j'=\{j|\Delta_j=max\{\Delta_i\}, j_{max}\leq i \leq k\}$;
    \RETURN $j'$
  \end{algorithmic}
\end{algorithm}
}

%The splits of the same $EC$ can be constructed by appending unique {\em salt} values to the plaintext value before encryption. In particular, to generate ${\varpi}$ splits of a given value $v$, we can concatenate $v$ with ${\varpi}$ unique values, and encrypt the ${\varpi}$ concatenated values by using the $AES$ encryption.
Essentially the splitting procedure can be implemented by the probabilistic encryption. In particular, for any plaintext value $p$, it is encrypted as $e=<r, F_k(r)\oplus p>$, where $r$ is a random string of length $\lambda$, $F$ is a pseudorandom function, $k$ is the key, and $\oplus$ is the {\em XOR} operation. The splits of the same $EC$ can easily be generated by using different random values $r$. 
In order to decrypt a ciphertext $e=<r, s>$, we can recover $p$ by calculating $p=F_k(r)\oplus s$. 
For security concerns, we require that the plaintext values that appear in different $ECGs$ are never encrypted as the same ciphertext value. The reason will be explained in Section \ref{sc:semi}. The complexity of the $S\&S$ step is $O(t^2)$, where $t$ is the number of equivalence classes of $D$. As $t=O(nq)$, the complexity is $O(n^2q^2)$, where $n$ is the number of tuples in $D$, and $q$ is the number of $MASs$.
%As $n$ dominates the $p$ {\bf what is p? $p=|ECG|$?}, which is the number of $ECs$ of $D$, the total complexity of Step 2 is $O(n)$. 

\vspace{-0.05in}
\subsection{Step 3: Conflict Resolution}
So far the grouping and splitting \& scaling steps are applied on one single $MAS$. In practice, it is possible that there exist multiple $MASs$. 
%In other words, there exist attribute sets $X, Y$ such that the stripped partition $\pi_{X}\neq \emptyset$, $\pi_{Y}\neq \emptyset$ and there is not a containment relationship between $X$ and $Y$. 
The problem is that, applying grouping and splitting \& scaling separately on each $MAS$ may lead to conflicts. 
%For instance, we may want to construct the duplicated value of tuple $T$ accoriding to one $MAS$ on attribute $A$, while keep $T$ free of scaling accoriding to another $MAS$ on attribute $B$. 
There are two possible scenarios of conflicts: \\
\noindent (1) {\em Type-1. Conflicts due to scaling}: there exist tuples that are required to be scaled by one $MAS$ but not so by another $MAS$; and \\
\noindent (2) {\em Type-2. Conflicts due to shared attributes}: there exist tuples whose value on attribute(s) $Z$ are encrypted differently by multiple $MASs$, where $Z$ is the overlap of these $MASs$. 

It initially seems true that there may exist the conflicts due to splitting too; some tuples are required to be split according to one $MAS$ but not by another. But our further analysis shows that such conflicts only exist for overlapping $MASs$, in particular, the type-2 conflicts. Therefore, dealing with type-2 conflicts covers the conflicts due to splitting. 

The aim of the conflict resolution step is to synchronize the encryption of all $MASs$. 
Given two $MASs$ of the attribute sets $X$ and $Y$, we say these two $MASs$ {\em overlap} if $X$ and $Y$ overlap at at least one attribute. Otherwise, we say the two $MASs$ are non-overlapping. 
Next, we discuss the details of the conflict resolution for non-overlapping $MASs$ (Section \ref{sc:nonoverlapping}) and overlapping $MASs$ (Section \ref{sc:inter}). 

\begin{figure*}[ht]
\begin{center}
\begin{small}
\begin{tabular}{cccccc}
\begin{tabular}{| c | c | c | c |}
  \hline
  ID & A & B & C \\ \hline
  $r_1$ & $a_3$ & $b_2$ & $c_1$ \\
  $r_2$ & $a_1$ & $b_2$ & $c_1$ \\
  $r_3$ & $a_2$ & $b_2$ & $c_1$ \\
  $r_4$ & $a_2$ & $b_2$ & $c_2$ \\
  $r_5$ & $a_3$ & $b_2$ & $c_2$ \\ 
  $r_6$ & $a_1$ & $b_1$ & $c_3$ \\ 
  \hline
\end{tabular}
& 
\begin{tabular}{| c | c | c |}
\hline
  ID & A & B \\ \hline
  $r_1$ & $\hat{a}_3^1$ & $\hat{b}_2^1$ \\
  $r_2$ & $\hat{a}_1^1$ & $\hat{b}_2^3$ \\
  $r_3$ & $\hat{a}_2^1$ & $\hat{b}_2^2$ \\
  $r_4$ & $\hat{a}_2^1$ & $\hat{b}_2^2$ \\
  $r_5$ & $\hat{a}_3^1$ & $\hat{b}_2^1$ \\ 
  $r_6$ & $\hat{a}_1^2$ & $\hat{b}_1^1$ \\ 
  \hline
\end{tabular}
&
\begin{tabular}{| c | c | c |}
\hline
  ID & B & C  \\ \hline
  $r_1$ & $\hat{b}_2^3$ & $\hat{c}_1^1$ \\  
  $r_2$ & $\hat{b}_2^3$ & $\hat{c}_1^1$ \\  
  $r_3$ & $\hat{b}_2^3$ & $\hat{c}_2^1$ \\  
  $r_4$ & $\hat{b}_2^4$ & $\hat{c}_2^1$ \\ 
  $r_5$ & $\hat{b}_2^4$ & $\hat{c}_2^1$ \\ 
  $r_6$ & $\hat{b}_1^1$ & $\hat{c}_3^1$ \\ 
  \hline
\end{tabular}
&
\begin{tabular}{| c | c | c | c |}
  \hline
   ID & A & B & C \\ \hline
  $r_1$ & $\hat{a}_3^1$ & $\hat{b}_2^1$ / $\hat{b}_2^3$ & $\hat{c}_1^1$ \\
  $r_2$ & $\hat{a}_1^1$ & $\hat{b}_2^3$ & $\hat{c}_1^1$ \\
  $r_3$ & $\hat{a}_2^1$ & $\hat{b}_2^2$ / $\hat{b}_2^3$ & $\hat{c}_1^1$ \\
  $r_4$ & $\hat{a}_2^1$ & $\hat{b}_2^2$ / $\hat{b}_2^4$ & $\hat{c}_2^1$ \\
  $r_5$ & $\hat{a}_3^1$ & $\hat{b}_2^1$ / $\hat{b}_2^4$ & $\hat{c}_2^1$ \\
  $r_6$ & $\hat{a}_1^2$ & $\hat{b}_1^1$ & $\hat{c}_3^1$ \\
  \hline
\end{tabular}
&
\begin{tabular}{| c | c | c | c |}
  \hline
   ID & A & B & C \\ \hline
  $r_1$ & $\hat{a}_3^1$ & $\hat{b}_2^1$ & $\hat{c}_1^1$ \\
  $r_2$ & $\hat{a}_1^1$ & $\hat{b}_2^3$ & $\hat{c}_1^1$ \\
  $r_3$ & $\hat{a}_2^1$ & $\hat{b}_2^3$ & $\hat{c}_1^1$ \\
  $r_4$ & $\hat{a}_2^1$ & $\hat{b}_2^4$ & $\hat{c}_2^1$ \\
  $r_5$ & $\hat{a}_3^1$ & $\hat{b}_2^1$ & $\hat{c}_2^1$ \\
  $r_6$ & $\hat{a}_1^2$ & $\hat{b}_1^1$ & $\hat{c}_3^1$ \\
  \hline
\end{tabular}
&
\begin{tabular}{| c | c | c | c |}
  \hline
  ID & A & B & C \\ \hline
  $r_1$ & $\hat{a}_3^1$ & $\hat{b}_2^1$ & $\hat{c}_1^2$ \\
  $r_2$ & $\hat{a}_1^1$ & $\hat{b}_2^3$ & $\hat{c}_1^1$ \\
  $r_3$ & $\hat{a}_2^1$ & $\hat{b}_2^2$ & $\hat{c}_1^3$ \\
  $r_4$ & $\hat{a}_2^1$ & $\hat{b}_2^2$ & $\hat{c}_2^2$ \\
  $r_5$ & $\hat{a}_3^1$ & $\hat{b}_2^1$ & $\hat{c}_2^3$ \\
  $r_6$ & $\hat{a}_1^2$ & $\hat{b}_1^1$ & $\hat{c}_3^1$ \\
  $r_7$ & $\hat{a}_3^2$ & $\hat{b}_2^3$ & $\hat{c}_1^1$ \\
  $r_8$ & $\hat{a}_2^2$ & $\hat{b}_2^3$ & $\hat{c}_1^1$ \\
  $r_9$ & $\hat{a}_2^3$ & $\hat{b}_2^4$ & $\hat{c}_2^1$ \\
  $r_{10}$ & $\hat{a}_3^3$ & $\hat{b}_2^4$ & $\hat{c}_2^1$ \\
  \hline
\end{tabular}
\\ 
(a) Dataset $D$
&
(b) $Enc_X(D)$
&
(c) $Enc_Y(D)$
&
(d) Encryption with 
&
(e) $\hat{D}_1$: Encryption by  
&
(f) $\hat{D}_2$: Encryption by 
\\
$F: C\rightarrow B$
& 
&
&
conflicts
&
the naive solution
&
conflict resolution
\\
\end{tabular}
\vspace{-.1in}
\caption{\label{fig:sync2} An example of conflict resolution of two overlapping $MASs$}
\end{small}
\end{center}
\vspace{-.3in}
\end{figure*}

\subsubsection{Non-overlapping $MASs$}
\label{sc:nonoverlapping}
Given two $MASs$ of attribute sets $X$ and $Y$ such that $X$ and $Y$ do not overlap, only the type-1 conflicts (i.e., conflicts due to scaling) are possible. WLOG we assume a tuple $r$ is required to be scaled by applying scaling on the $ECG$ of $r_X$ but not on any $ECG$ of $r_Y$. The challenge is that simply scaling of $r$ makes the $ECG$ of $r_Y$ fails to have homogenized frequency anymore. To handle this type of conflicts, first, we execute the grouping and splitting \& scaling over $\pi_X$ and $\pi_Y$ independently. Next, we look for any tuple $r$ such that $r$ is required to have $\ell$ ($\ell>1$) copies to be inserted by splitting \& scaling over $\pi_X$ but free of split and scaling over $\pi_Y$. 
 %$T\in C_i \in ECG_m \subset \pi_X$. According to the split\& scaling scheme over $ECG_m$, we may want to duplicate $r$. At the same time, $r\in C_j \in ECG_n \subset \pi_Y$. While the frequency distribution on $ECG_n$ decides that $r$ should not be split or scaled. Now we are in a dilemma that we need to duplicate $r[X]$ but not $r[Y]$. As a consequence, we should not simply add a duplication of $r$ as a fake record. Instead, 
 For such $r$, we modify the values of the $\ell$ copies of tuple $r$, ensuring that for each copy $r'$, $r'[X]=r[X]$ while $r'[Y]\neq r[Y]$. 
 %We require that each copy $r'$ has a unique $r'[Y]$ value. 
 To avoid collision between $ECGs$, we require that no $r'[Y]$ value exists in the original dataset $D$. By doing this, we ensure that the $ECGs$ of both $r_X$ and $r_Y$ still achieve the homogenized frequency distribution. Furthermore, by assigning new and unique values to each copy on the attribute set $Y$, 
 we ensure that $MASs$ are well preserved (i.e., both $MASs$ $X$ and $Y$ do not change to be $X\cup Y$). The complexity of dealing with type-1 conflicts is $O(n'q)$, where $n'$ is the number of the tuples that have conflicts due to scaling, and $q$ is the number of $MASs$.
  %(2) We do not change the frequency distribution in $ECG_n$ because we add a collision-free value at $r'[Y]$, which means that the equivalence class containing the value in $r[Y]'$ only includes one record, which is $r'$. \\
  %(3) We do not change the functional dependency or $Y$. Because $Y$ is a maximal attribute set, there may exist a functional dependency in $Y$. The value in $r'[Y]$ does not have any collision with any other value in $Y$, so if there is a functional dependency, it is not affected by $r'$.

\subsubsection{Overlapping $MASs$}
\label{sc:inter}
%Because $X$ and $Y$ have attribute intersection, and at the same time we need to enforce the principle that the same plaintext value must not share the same ciphertext value in different groups, we should be careful with the encryption on $X$ and $Y$. 
When $MASs$ overlap, both types of conflicts are possible. 
The type-1 conflicts can be handled in the same way as for non-overlapping $MASs$ (Section \ref{sc:nonoverlapping}). In the following discussion, we mainly focus on how to deal with the type-2 conflicts (i.e., the conflicts due to shared attributes). We start our discussion from two overlapping $MASs$. Then we extend to the case of more than two overlapping $MASs$.

\noindent{\bf Two overlapping MASs.} 
We say two $ECs$ $C_i \in \pi_X$ and $C_j \in \pi_Y$ are \emph{conflicting} if $C_i$ and $C_j$ share at least one tuple. 
We have the following theorem to show that conflicting $ECs$ never share more than one tuple.  
\vspace{-.05in}
\begin{theorem}
\label{th:inter}
  Given two overlapping $MASs$ $X$ and $Y$, for any pair of $ECs$ $C_i \in \pi_X$ and $C_j \in \pi_Y$, $|C_i \cap C_j| \leq 1$. 
\end{theorem}

The correctness of Theorem \ref{th:inter} is straightforward: if $|C_i \cap C_j| > 1$, there must exist at least one equivalence class of the partition $\pi_{X \cup Y}$ whose size is greater than 1. Then $X\cup Y$ should be a $MAS$ instead of $X$ and $Y$. Theorem \ref{th:inter} ensures the efficiency of the conflict resolution, as it does not need to handle a large number of tuples. 

A naive method to fix type-2 conflicts is to assign the same ciphertext value to the shared attributes of the two conflicting $ECs$. As an example, consider the table $D$ in Figure \ref{fig:sync2} (a) that consists of two $MASs$: $X=\{A,B\}$ and $Y=\{B,C\}$. Figure \ref{fig:sync2} (b) and Figure \ref{fig:sync2} (c) show the encryption $Enc_X(D)$ over $X$ and $Enc_Y(D)$ over $Y$ independently. The conflict appears at tuples $r_1$, $r_3$, $r_4$, and $r_5$ on attribute $B$ (shown in Figure \ref{fig:sync2} (d)). Following the naive solution, only one value is picked for tuples $r_1$, $r_3$, $r_4$, and $r_5$ on attribute $B$. Figure \ref{fig:sync2} (e) shows a conflict resolution scheme $\hat{D}_1$ by the naive method. 
%This scheme has several weakness. 
This scheme is incorrect as the FD $F: C\rightarrow B$ in $D$ does not hold in $\hat{D}_1$ anymore. 

We design a robust method to resolve the type-2 conflicts for two overlapping $MASs$. 
Given two overlapping $MASs$ $X$ and $Y$, let $Z = X \cap Y$, for any tuple $r$, let $r^X[Z]$ and $r^Y[Z]$ ($r^X[Z] \neq r^Y[Z]$) be the value constructed by encryption over $X$ and $Y$ independently. We use $X-Z$ ($Y-Z$, resp.) to denote the attributes that appear in $X$ ($Y$, resp.) but not $Z$. Then we construct two tuples $r_1$ and $r_2$: 
\begin{ditemize}
\item $r_1$: $r_1[X-Z] = r[X-Z]$, $r_1[Y - Z ] = v_X$, and $r_1[Z] = r^X[Z]$;
\item $r_2$: $r_2[X-Z] = v_Y$, $r_2[Y - Z ] = r[Y-Z]$, and $r_2[Z] = r^Y[Z]$.  
\end{ditemize}
where $v_X$ and $v_Y$ are two values that do not exist in $D$. Note that both $X-Z$ and $Y-Z$ can be sets of attributes, thus $v_X$ and $v_Y$ can be set of values. 
Tuples $r_1$ and $r_2$ replace tuple $r$ in $\hat{D}$. 
As an example, consider the table in Figure \ref{fig:sync2} (a), the conflict resolution scheme by following our method is shown in Figure \ref{fig:sync2} (f). For example, $r_1$ and $r_7$ in Figure \ref{fig:sync2} (f) are the two records constructed for the conflict resolution of $r_1$ in Figure \ref{fig:sync2} (d). 
%\in C_i \in ECG_m \subset \pi_X$ and $r \in C_j \in ECG_n \subset \pi_Y$, if we want to assign value $v_1$ to $r[Z]$ for $Z = X\cap Y$ in $Enc_X(D)$, and assign value $v_2$ to $r[Z]$ in $Enc_Y(D)$, we should prepare two records $r_1$ and $r_2$ with $r_1[X]=Enc_X(r[X])$ and $r_1[R-X]$ having collision-free values, and $r_2[Y]=Enc_Y(r[Y])$ and $r_2[R-Y]$ having collision-free values. We need to set $r=r_1$ and append $r_2$ to the encrypted dataset. In Figure \ref{fig:sync2}(e), we assign $r_1, r_3, r_4, r_5$ new values and add $r_6, r_7, r_8, r_9$. With this method, we also ensures that: \\

Our conflict resolution method guarantees that the ciphertext values of each $ECG$ of $X$ and $Y$ are of homogenized frequency. However, it requires to add additional records. Next, we show that the number of records added by resolution of both types of conflicts is bounded.
\vspace{-0.05in}
\begin{theorem}
\label{theorem:bound1}
Given a dataset $D$, let $\hat{D}_1$ be the dataset after applying grouping and splitting \& scaling on $D$, and $\hat{D}_2$ be the dataset after conflict resolution on $\hat{D}_1$, then 
 $|\hat{D}_2| - |\hat{D}_1| \leq hn$, where $h$ is the number of overlapping $MAS$ pairs, and $n$ is the size of $D$.
\end{theorem}
\vspace{-0.05in}
Due to the space limit, we only give the proof sketch here. Note that the resolution of type-1 conflicts does not add any fake record. Thus we only prove the bound of the new records for resolution of type-2 conflicts. This type of conflicts is resolved by replacing any conflicting tuple with two tuples for each pair of conflicting $ECs$. Since two overlapping $MASs$ have $n$ conflicting equivalence class pairs at most, there will be at most $hn$ new records inserted for this type of resolution for $h$ overlapping $MAS$ pairs. We must note that $hn$ is a loose bound. In practice, the conflicting $ECs$ share a  small number of tuples. The number of such tuples is much smaller than $n$. We also note that when $h$ = 0 (i.e., there is no overlapping $MAS$), no new record will be inserted. 

\begin{figure*}[t!]
\begin{center}
\begin{small}
\begin{tabular}{ccc}
\begin{tabular}{|l|l|l|l|l|}
\hline
&EC&A&B&Freq\\
\hline
\multirow{2}{*}{$ECG_1$} &$C_1$ & $a_1$ & $b_1$ & 5 \\\hhline{~----}
						             &$C_2$ & $a_2$ & $b_3$ & 2 \\\hline
\multirow{2}{*}{$ECG_2$} &$C_3$ & $a_1$ & $b_2$ & 4\\\hhline{~----}
						             &$C_4$ & $a_2$ & $b_4$ & 3\\\hline 
\end{tabular}
&
\begin{tabular}{|l|l|l|l|}
\hline
&A&B&Freq\\
\hline
\multirow{3}{*}{$ECG_1$} & $\hat{a}_1^1$ & $\hat{b}_1^1$ & 3 \\\hhline{~---}
						&  $\hat{a}_1^2$ & $\hat{b}_1^2$ & 3 \\\hhline{~---}
						& $\hat{a}_2^1$ & $\hat{b}_3^1$ & 3 \\\hline
\multirow{2}{*}{$ECG_2$} & $\hat{a}_1^3$ & $\hat{b}_2^1$ & 4 \\\hhline{~---}
						& $\hat{a}_2^2$ & $\hat{b}_4^1$ & 4 \\\hline
\end{tabular}
&
\begin{tabular}{|l|l|l|}
\hline
A&B&Freq\\
\hline
$\hat{a}^3$ & $\hat{b}^5$ & 1 \\\hline
$\hat{a}^3$ & $\hat{b}^6$ & 1 \\\hline
$\hat{a}^4$ & $\hat{b}^7$ & 1 \\\hline
$\hat{a}^4$ & $\hat{b}^8$ & 1 \\\hline
$\hat{a}^5$ & $\hat{b}^9$ & 1 \\\hline
$\hat{a}^5$ & $\hat{b}^{10}$ & 1 \\\hline
\end{tabular}
\\
(a) Base table $D$ &
(b) $\hat{D}$: encryption by Step 1 - 3 &
(c) Constructed $\Delta D$ to remove 
\\
($A\rightarrow B$ does not hold)&
($A\rightarrow B$ becomes false positive) &
false positive $FD$ $A\rightarrow B$
\end{tabular}
\vspace{-0.1in}
\caption{\label{table:ss} An example of eliminating false positive $FDs$}
\end{small}
\end{center}
\vspace{-0.1in}
%\end{figure*}
\end{figure*}

\noindent{\bf More than Two Overlapping MASs.} 
When there are more than two overlapping $MASs$, one way is to execute the encryption scheme that deals with two overlapping $MASs$ repeatedly for every two overlapping $MASs$. This raises the question in which order the  $MASs$ should be processed. We have the following theorem to show that indeed the conflict resolution is insensitive to the order of $MASs$.
\vspace{-0.05in}
\begin{theorem}
\label{th:order}
Given a dataset $D$ that contains a set of overlapping $MASs$ $M$, let $\hat{D}_1$ and $\hat{D}_2$ be the datasets after executing conflict resolution in two different orders of $M$, then $|\hat{D}_1|=|\hat{D}_2|$.
\end{theorem}
\begin{proof}
It is easy to show that no new record is inserted by resolution of type-1 conflicts. To fix type-2 conflicts, for each pair of overlapping $MAS$s, every conflicting tuple is replaced by two tuples. The number of conflicting tuples is equal to the number of conflicting equivalence class pairs of the two overlapping $MAS$s. Assume there are $q$ such overlapping $MAS$ pairs, each having $o_i$ pairs of conflicting equivalence class pairs. The conflict resolution method adds $O=\sum\limits_{i=1}^q o_i$ records in total. The number of records is independent from the orders of $MAS$s in which conflict resolution is executed. 
\end{proof}
\vspace{-0.05in}
Since the order of $MASs$ does not affect the encryption, we pick the overlapping $MAS$ pairs randomly. For each overlapping $MAS$ pair, we apply the encryption scheme for two overlapping $MASs$. We repeat this procedure until all $MASs$ are processed. 

\nop{ 
We have the following theorem to show that the three steps of $F^2$ preserves all $FDs$ in $D$.
\vspace{-0.05in}
\begin{theorem}
\label{theorem:fd1}
{\em
Given the dataset $D$, let $\hat{D}$ be the dataset after applying grouping, splitting \& scaling and conflict resolution, then all FDs that hold on $D$ still hold on $\hat{D}$.   
}
\end{theorem}
\vspace{-0.05in}
The proof of Theorem \ref{theorem:fd1} can be found in our full paper \cite{vldb2016full}. 
}

\subsection{Step 4. Eliminating False Positive FDs}
\label{sc:fdfp}

Before we discuss the details of this step, we first define false positive FDs. Given the dataset $D$ and its encrypted version $\hat{D}$, we say the FD $F$ is a {\em false positive} if $F$ does not hold in $D$ but holds in $\hat{D}$. We observe that the encryption scheme constructed by Step 1 - 3 may lead to false positive FDs. Next, we show an example of false positive FDs. 
%, no splits of any two $ECs$ have collision, even for the original $ECs$ that have collisions. This may 
\vspace{-0.05in}
\begin{example}
\label{exp:1}
Consider the table $D$ in Figure \ref{table:ss} (a). Apparently the $FD$ $F:A \rightarrow B$ does not exist in $D$, as the two $ECs$ $C_1$ and $C_3$ have collisions. However, in the encrypted dataset $\hat{D}$ that is constructed by Step 1 - 3 (Figure \ref{table:ss} (b)), since no splits of any two $ECs$ have collision anymore, $F:A \rightarrow B$ now holds in $\hat{D}$. 
%The reason is obvious; for the two $ECs$ $E_1$ and $E_3$ in $D$ that have collisions (which fails the $F: A\rightarrow B$ in $D$), . 
\end{example}
\vspace{-0.05in}

Indeed, we have the following theorem to show that free of collision among $EC$s is the necessary and sufficient conditions for the existence of FDs. 
\vspace{-.05in}
\begin{theorem}
\label{theorem:fd}
For any MAS $X$, there must exist a FD on $X$ if and only if for any two $ECs$ $C_i$ and $C_j$ of $\pi_X$, $C_i$ and $C_j$ do not have collision. 
\end{theorem}
\vspace{-.05in}
Following Theorem \ref{theorem:fd}, the fact that Step 1 - 3 construct only collision-free $ECs$ may lead to a large number of false positive $FDs$, which hurts the accuracy of dependency discovering by the server.

\begin{figure}[t!]
\begin{center}
\vspace{-0.05in}
\includegraphics[width=0.45\textwidth]{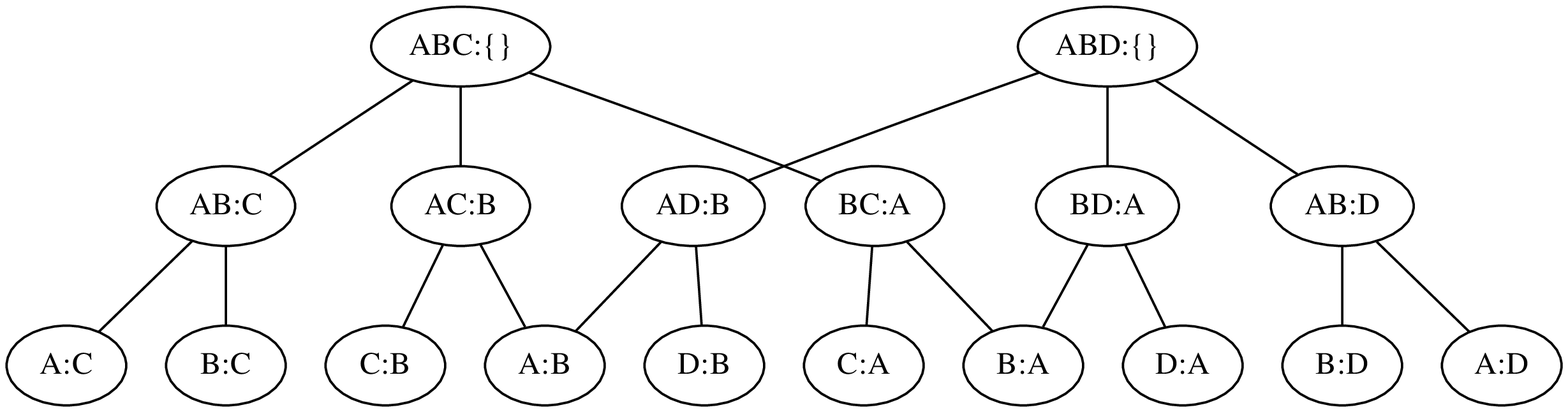}
\vspace{-0.1in}
\caption{\label{fig:fd_lattice}An example of FD lattice}
\end{center}
\vspace{-0.3in}
\end{figure}

The key idea to eliminate the false positive $FDs$ is to restore the collision within the $ECs$. Note that eliminating the false positive FD $F: X\rightarrow Y$ naturally leads to the elimination of all false positive FDs $F': X'\rightarrow Y$ such that $X'\subseteq X$. Therefore, we only consider eliminating the {\em maximum false positive FDs}  whose LHS is not a subset of LHS of any other false positive FDs. 

A simple approach to eliminate the false positive FDs is that the data owner informs the server the ECs with collisions. Though correct, this solution may bring security leakage as it implies which ciphertext values are indeed encrypted from the same plaintext.  For example,  consider Figure \ref{table:ss} (b). If the server is informed that $ECG_1$ and $ECG_2$ have collisions before encryption, it can infer that the three distinct ciphertext values in $ECG_1$ must only map to two plaintext values (given the two distinct ciphertext values in $ECG_2$). Therefore, we take a different approach based on adding artificial records to eliminate false-positive FDs. 

We use the {\em FD lattice} to help restore the $ECs$ with collision and eliminate false positive FDs. The lattice is constructed in a top-down fashion. Each $MAS$ corresponds to a level-1 node in the lattice. We denote it in the format $M: \{\}$, where $M$ is the $MAS$ that the node corresponds to. The level-2 nodes in the lattice are constructed as following. For each level-1 node $N$, it has a set of children nodes (i.e., level-2 nodes) of format $X: Y$, where $Y$ corresponds to a single attribute of $D$, and $X = M - \{Y\}$, where $M$ is the $MAS$ that node $N$ corresponds to. 
Starting from level 2, for each node $X: Y$ at level $\ell$ ($\ell \geq 2$), it has a set of children nodes of format  $X': Y'$, where $Y' = Y$, and $X' \subset X$, with $|X|' = |X| - 1$ (i.e., $X'$ is the largest subset of $X$).  Each  node at the bottom of the lattice is of format $X: Y$, where both $X$ and $Y$ are 1-attribute sets. An example of the FD lattice is shown in Figure \ref{fig:fd_lattice}.
 
Based on the FD lattice, the data owner eliminates the false positive FDs by the following procedure. Initially all nodes in the lattice are marked as ``un-checked''. Starting from the second level of lattice, for each node $N$ (in the format $X: Y$), the data owner checks whether there exists at least two $ECs$ $C_i, C_j\in\pi_{M}$ such that $C_i[X]=C_j[X]$ but $C_i[Y]\neq C_j[Y]$, where $M$ is the MAS that $N$'s parent node corresponds to. %Note here we only consider one single attribute in $A$, as all FDs only have 1-attribute RHS. %And we also only consider $X = M - \{Y\}$, instead of all subsets of $M - \{Y\}$, because if $X \not\rightarrow Y$, then it must be true that for any subset $X'\subseteq X$, $X'\not\rightarrow Y$. 
If it does, then the data owner does the following two steps. First, the data owner inserts %$k=\lceiling\frac{1}{\alpha}\rceiling$ 
$k=\lceil\frac{1}{\alpha}\rceil$ artificial record pairs $\{P_1, \dots, P_k\}$, where $\alpha$ is the given threshold for $\alpha$-security. The reason why we insert $k$ such pairs will be explained in Section \ref{sc:security}. Each $P_i$ consists of two records $r_i^1$ and $r_i^2$:
\vspace{-0.05in}
\begin{ditemize}
  \item $r_i^1:$ $r_i^1[X]=x_i$, $r_i^1[Y]=a_i^1$, and $r_i^1[MAS-X-\{Y\}]=v_i^1$;
  \item $r_i^2:$ $r_i^2[X]=x_i$, $r_i^2[Y]=a_i^2$, and $r_i^2[MAS-X-\{Y\}]=v_i^2$.
\end{ditemize}
\vspace{-0.05in}
where $x_i$, $a_i^1, a_i^2, v_i^1$ and $v_i^2$ are artificial values that do not exist in $\hat{D}$ constructed by the previous steps. We require that $a_i^1 \neq a_i^2$, and $v_i^1 \neq v_i^2$. We also require that all artificial records are of frequency one. Second, the data owner marks the current node and all of its descendants in the lattice as ``checked'' (i.e., the current node is identified as a maximum false positive FD). 
Otherwise (i.e., the $ECs$ do not have collisions), the data owner simply marks the current node as ``checked''. After checking all the nodes at the level $\ell$, the data owner moves to the level $\ell + 1$ and applies the above operation on the ``un-checked'' lattice nodes. The data owner repeats the procedure until all nodes in the lattice are marked as ``checked''.

Let $\Delta{D}$ be the artificial records that are constructed by the above iterative procedure. It is easy to see  that for any FD $X\rightarrow Y$ that {\em does not} hold in $D$, there always exist two $ECs$ $C_i'$ and $C_j'$ in $\Delta{D}$, where $C_i'[X]=C_j'[X]$ but $C_i'[Y]\neq C_j'[Y]$. This makes the FD $X\rightarrow Y$ that does not hold in $D$ fails to hold in $\hat{D}$ too. To continue our Example \ref{exp:1}, we show how to construct $\Delta D$. It is straightforward that the $MAS$ $\{A, B\}$ contains the $ECs$ that have collision (e.g. $E_1$ and $E_3$ in Figure \ref{table:ss} (a)). Assume $\alpha = 1/3$. Then $\Delta{D}$ (shown in Figure \ref{table:ss} (c)) consists of three pairs of artificial tuples, each pair consisting of two records of the same value on attribute $A$ but not on $B$. The false positive FD $F: A\rightarrow B$ does not exist in $\hat{D} + \Delta{D}$ any more. 

%It is worth noting that the complexity to search for $FD$ false positives within $MAS$ is much cheaper than $FD$ discovery. This is because to detect the non-existence of $F: X \rightarrow A$, the client only needs to search for two $ECs$ $C_i$ and $C_j$ with same value on $X$ but different value on $A$ in the partition of $MAS$, rather than comparing the partitions on $X$ and $X\cup\{A\}$. By building a map between the values on $X$ and $A$, the client can finish the computation by one scan of $\Pi_{MAS}$. So the time complexity of our method to remove $FD$ false positives is $O(2^{|MAS|}p)$, where $|MAS|$ is the number of attributes in $MAS$, and $p$ is the number of equivance classes on $MAS$. This method requires to build aother dataset $\Delta D$. The size of $\Delta D$ is bounded by $2^{|MAS|}2k$. Considering that in practice, $|MAS|$ is usually a small number, the number of added records is very small compared to the size of original dataset. 
\nop{
{\bf
In the existence of multiple $MAS$s, if we apply the false positive FD removal method separately on every $MAS$, we may have to check the validity of some FDs multiple times. For example, given $MAS_1=\{A, B, C, D\}$ and $MAS_2=\{A, B, C, E\}$, on both of them, we need to do the checking on $\{AB\} \rightarrow C$, $\{AC\} \rightarrow B$ and $\{BC\} \rightarrow A$. To avoid it, we can follow the level-wise BFS search logic as \cite{agrawal1994fast}. Assume the maximum size of all the $MAS$s is $\ell$. Initially we generate the candidate FDs of length $\ell$ from the set of $MAS$s, check their validity and generate the candidate FDs of length $\ell-1$ from valid FDs of length $\ell$. We repeat this operation until we can find no candidate FDs. The difference from \cite{agrawal1994fast} is that our algorithm goes in the top-down direction, i.e., we generate candidate FDs of length $l-1$ from FDs of length $l$. In this way, we ensure that each candidate FD $X \rightarrow A$ is processed only once. 
}
}

Next, we show that the number of artificial records added by Step 4 is bounded.
\vspace{-0.05in}
\begin{theorem}
\label{theorem:bound}
Given a dataset $D$, let $M_1, \dots, M_q$ be the $MASs$ of $D$. Let $\hat{D}_1$ and $\hat{D}_2$ be the dataset before and after eliminating false positive FDs respectively. Then  
 \[2k \leq |\hat{D}_2| - |\hat{D}_1| \leq min(2km\binom{m-1}{[\frac{m-1}{2}]}, 2k\sum_{i=1}^{q} |M_i|\binom{|M_i|-1}{[\frac{|M_i|-1}{2}]}),\] where $k=\lceil\frac{1}{\alpha}\rceil$ ($\alpha$ as the given threshold for $\alpha$-security), $m$ is the number of attributes of $D$, $q$ is the number of $MASs$, and $|M_i|$ as the number of attributes in $M_i$.
\end{theorem}
\begin{proof}
Due to the limited space, we show the proof sketch here. 
We consider two extreme cases for the lower bound and upper bound of the number of artificial records. First, for the lower bound case, there is only one $MAS$ whose $ECs$ have collisions. It is straightforward that our construction procedure constructs $2k$ artificial records, where $k=\lceil\frac{1}{\alpha}\rceil$. Second, for the upper bound case, it is easy to infer that the maximum number of nodes in the FD lattice that needs to construct artificial records is $\sum_{i=1}^{q} |M_i|\binom{|M_i|-1}{[\frac{|M_i|-1}{2}]}$. For each such lattice node, there are $2k$ artificial records. So the total number of artificial records is no larger than $2k\sum_{i=1}^{q} |M_i|\binom{|M_i|-1}{[\frac{|M_i|-1}{2}]}$. On the other hand, as for each maximum false positive FD, its elimination  needs $2k$ artificial records to be constructed. Therefore, the total number of artificial records added by Step 4 equals $2ku$, where $u$ is the number of maximum false positive FDs. It can be inferred that $u \leq m\binom{m-1}{[\frac{m-1}{2}]}$. Therefore, the number of artificial records cannot exceed $min(2k\sum_{i=1}^{q} |M_i|\binom{|M_i|-1}{[\frac{|M_i|-1}{2}]}, 2km\binom{m-1}{[\frac{m-1}{2}]})$.  Note that the number of artificial records is independent of the size of the original dataset $D$. It only relies on the number of attributes of $D$ and $\alpha$. 
\end{proof}
\vspace{-0.05in}

The time complexity of eliminating false positive $FDs$ for a single $MAS$ $M$ is $O(2^{|M|}t)$, where $|M|$ is the number of attributes in $M$, and $t$ is the number of $ECs$ of $M$.
In our experiments, we observe $t<<n$, where $n$ is the number of records in $D$. 
For instance, on a benchmark dataset of $n=15,000,000$, the average value of $t$ is $11,828$.
Also, the experiment results on all three datasets show that at most $|M|=\frac{m}{2}$.
With the existence of $q>1$ $MAS$s, the time complexity is $O(\sum_{i=1}^{q}{2^{|M_i|}t_i})$, where $t_i$ is the number of equivalence classes in $M_i$, and $|M_i|$ is the number of attributes of $MAS$ $M_i$. 
Considering that $t_i<<n$, the total complexity is comparable to $O(nm^2)$.
%As $|M|$ is bounded by the number of the attributes of $D$, the number of added artificial records relies on the schema of $D$, not the instance size of $D$. 
%As the number of the attributes of $D$ is always much smaller than the size of $D$,  the artificial records always takes a small portion of the dataset. 

We have the following theorem to show that the FDs in $\hat{D}$ and $D$ are the same by our 4-step encryption. 
\vspace{-0.05in}
\begin{theorem}
\label{theorem:fd2}
Given the dataset $D$, let $\hat{D}$ be the dataset after applying Step 1 - 4 of $F^2$ on $D$, then: (1) any FD of $D$ also hold on $\hat{D}$; and (2) any FD $F$ that does not hold in $D$ does not hold in $\hat{D}$ either. 
\end{theorem}
\begin{proof}
First, we prove that the grouping, splitting \& scaling and conflict resolution steps keep the original FDs. We prove that for any FD $X\rightarrow A$ that holds on $D$, it must also hold on $\hat{D}$. 
We say that a partition $\pi$ is a {\em refinement} of another partition $\pi'$ if every equivalence class in $\pi$ is a subset of some equivalence class ($EC$) of $\pi'$. It has been proven that the functional dependency $X \rightarrow A$ holds if and only if $\pi_X$ refines $\pi_{\{A\}}$\cite{huhtala1999tane}. For any functional dependency $X \rightarrow A$, we can find a $MAS$ $M$ such that $(X\cup \{A\}) \subset M$. Obviously, $\pi_M$ refines $\pi_X$, which means for any $EC$ $C \in \pi_M$, there is an $EC$ $C_p \in \pi_X$ such that $C \subset C_p$. Similarly, $\pi_M$ refines $\pi_{\{A\}}$. So for any equivalence class $C \in \pi_M$, we can find $C_p \in \pi_X$ and $C_q \in \pi_{\{A\}}$ such that $C \subset C_p \subset C_q$.
First, we prove that our grouping over $M$ keeps the FD: $X \rightarrow A$. It is straightforward that grouping $ECs$ together does not affect the FD. The interesting part is that we add fake $ECs$ to increase the size of a $ECG$. Assume we add a fake equivalence class $C_f$ into $\pi_M$. Because $C_f$ is {\em non-collisional}, $\pi_X = \pi_X \cup \{C_f\}$ and $\pi_{\{A\}} = \pi_{\{A\}} \cup \{C_f\}$. Therefore, $\pi_X$ still refines $\pi_{\{A\}}$. The FD is preserved.
Second, we show that our splitting scheme does not break the FD: $X \rightarrow A$. Assume that we split the equivalence class $C \in \pi_M$ into $\varpi$ unique equivalence classes $C^1, \dots, C^{\varpi}$. After the split, $C_p = C_p - C$, $C_q = C_q - C$. This is because the split copies have unique ciphertext values. As a result, $\pi_X = \pi_X \cup \{C^1, \dots, C^{\varpi}\}$ and $\pi_{\{A\}} = \pi_{\{A\}} \cup \{C^1, \dots, C^{\varpi}\}$. It is easy to see that $\pi_X$ is still a refinement of $\pi_{\{A\}}$. The scaling step after splitting still preserves the FD, as it only increases the size of the equivalence class $C \in \pi_M$ by adding additional copies. The same change applies to both $C_p$ and $C_q$. So $C_p$ is still a subset of $C_q$. As a result, the FD: $X \rightarrow A$ is preserved after splitting and scaling.
Lastly, we prove that our conflict resolution step keeps the $FD$s. First, the way of handling non-overlapping $MASs$ is FD-preserving because we increase the size of an equivalence class in a partition while keeping the stripped partitions of the other $MASs$. Second, the conflict resolution for overlapping $MASs$ is also FD-preserving. Assume $C \in \pi_M$ and $C' \in \pi_{N}$ conflict over a tuple $r$. According to our scheme, we use $r_1$ and $r_2$ to replace $r$ with $r_1[M]=r^M[M]$, $r_2[N]=r^N[N]$, $r_1[N-M]$ and $r_2[M-N]$ having new values. The effect is to replace $r \in C$ with $r_1$. This change does not affect the fact that $C_p \subset C_q$. Therefore the $FDs$ are still preserved. 
Here we prove that by inserting artificial records, all original FDs are still kept while all the false positive FDs are removed. 

Next, we prove that insertion of artificial records preserves all original FDs. The condition to insert fake records is that there exists equivalence classes $C_i$ and $C_j$ such that $C_i[X]=C_j[X]$ but $C_i[Y] \neq C_j[Y]$ on the attribute sets $X$ and $Y$. For any FD that holds on $D$, this condition is never met. Hence the real FDs in $D$ will be kept.

Last, we prove that any FD $F: X \rightarrow Y$ that does not hold in $D$ is also not valid in $\hat{D}$.
First we show that if there does not exist a $MAS$ $M$ such that $X\cup \{Y\} \not\subset M$, $X \rightarrow Y$ can not hold in $\hat{D}$. Since our splitting \& scaling procedure ensures that  different plaintext values have different ciphertext values, and each value in any equivalence class of a $MAS$ is encrypted to a unique ciphertext value, the set of $MAS$s in $D$ must be the same as the set of $MAS$s in $\hat{D}$. As a consequence, in $\hat{D}$,  there do not exist any two records $r_i$, $r_j$ such that $r_i[X]=r_j[X]$ and $r_i[Y]=r_j[Y]$. Therefore, $X \rightarrow Y$ cannot be a FD in $\hat{D}$. Second, we prove that for any $MAS$ $M$ such that  $X\cup \{Y\} \subset M$,  if $F$ does not hold on $D$, then $F$ must not hold on $\hat{D}$. 
\end{proof}
\vspace{-0.05in}

\vspace{-.05in}
\section{Security Analysis}
\label{sc:security}

In this section, we analyze the security guarantee of $F^2$ against the frequency analysis attack, for both cases of without and under Kerckhoffs's principle. We assume that the attacker can be the compromised server. 
\nop{
%It may launch two types of attack, based on the various adversary knowledge it has. In both attacks, {\bf Freq}  refers to the exact frequency information of the original dataset $D$. 
%\item {\bf Freq+FD}: With the knowledge of the exact frequency information of $D$, together with the $FDs$ that can be  discovered from $\hat{D}$, the attacker may try to re-construct $D$. 

{\bf (1) Freq+$\hat{D}$}: The attacker can launch the {\em frequency  analysis attack}; in particular, he can match the ciphertext values in $\hat{D}$ with the plaintext values in $D$ by their frequency. 

{\bf (2) Freq+$F^2$}: The attacker knows the details of the $F^2$ encryption algorithm, and tries to break the cipher by utilizing the knowledge of the scheme together with the frequency distribution information of $D$. 
}
\nop{
Table \ref{table:attackoverview} summarizes the various adversary knowledge for these three different attacks. 
\begin{table*} [ht!]
\begin{small}
\begin{center}
\begin{tabular}{|c|c|c|c|c|}
\hline
\multicolumn{4}{|c|}{Adversary Knowledge} & \multirow{2}{*}{Attack} \\\cline{1-4}
  Plaintext frequency & FD & Ciphertext frequency & Encryption scheme &  \\\hline
  Y & Y & N & N & Plaintext-only attack\\\hline
  Y & Y & Y & N & Known-ciphertext attack\\\hline
  Y & Y & Y & Y & Known-scheme attack\\\hline
\end{tabular}
\caption{\label{table:attackoverview}\small{Summary of attacks.}}
\end{center}
\end{small}
\vspace{-0.3in}
\end{table*}
}

\nop{
\subsection{Freq+FD Attack} 

\begin{figure}[ht]
\begin{center}
\begin{small}
\begin{tabular}{cc}
\begin{tabular}{| c | c |}
\hline
A & B \\
\hline
$a_1$: 10 & $b_1$: 16\\\hline
$a_2$: 6 & $b_2$: 5\\\hline
$a_3$: 2 &  \\\hline
$a_4$: 3 &  \\\hline
\end{tabular}
&
\begin{tabular}{|c|c|}
\hline
A & B \\
\hline
$a_1$: 10 & $b_1$: 15\\\hline
$a_2$: 5 & $b_2$: 5\\\hline
$a_3$: 2 & \\\hline
$a_4$: 3 & \\\hline
\end{tabular}
\\
(a) Freq. of $D_1$ ($F: A\rightarrow B$)
&
(b) Freq. of $D_2$ ($F: A\rightarrow B$)
\end{tabular}
\\
\begin{tabular}{cc}
\includegraphics[width=0.13\textwidth]{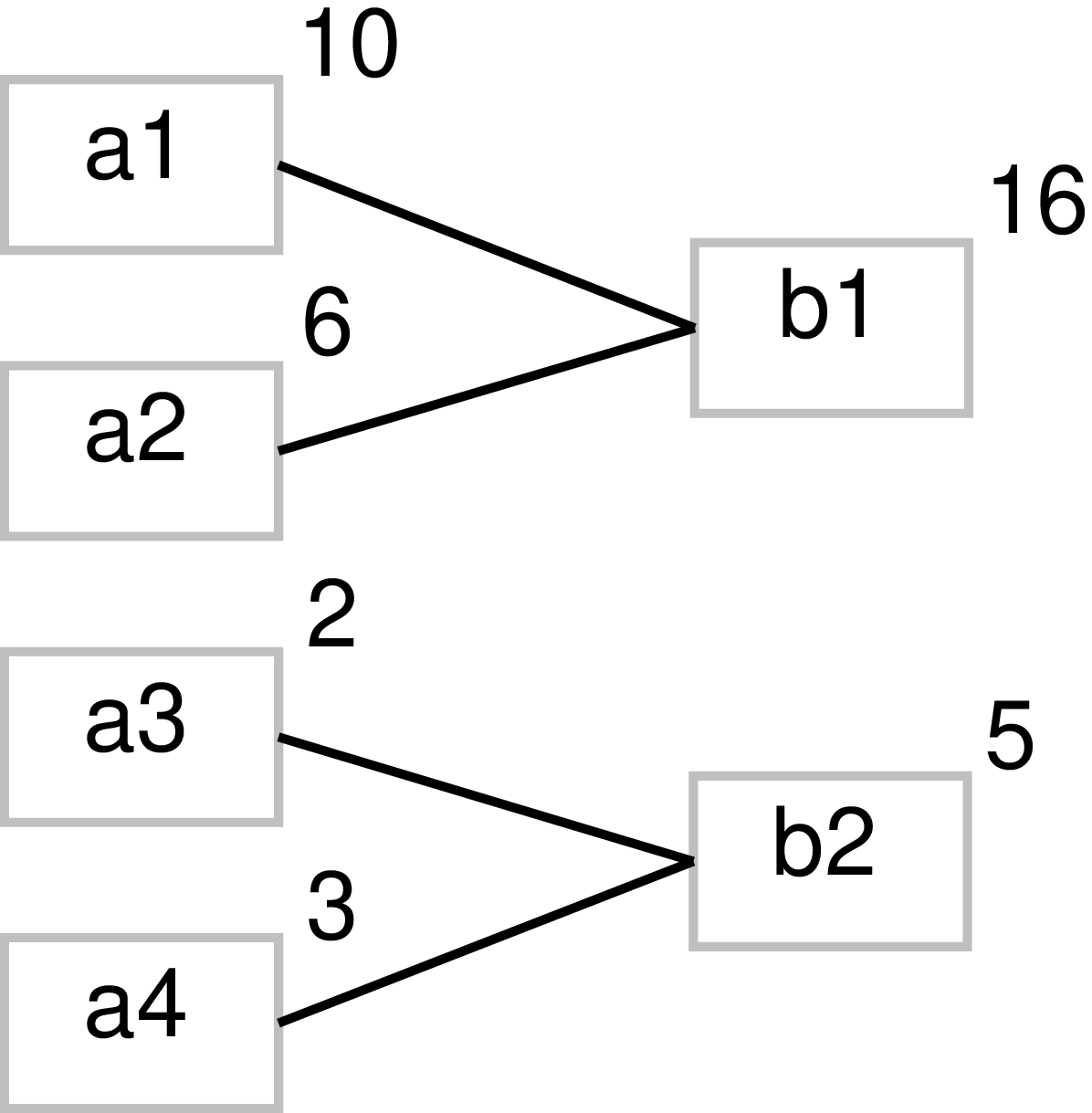}
&
\includegraphics[width=0.26\textwidth]{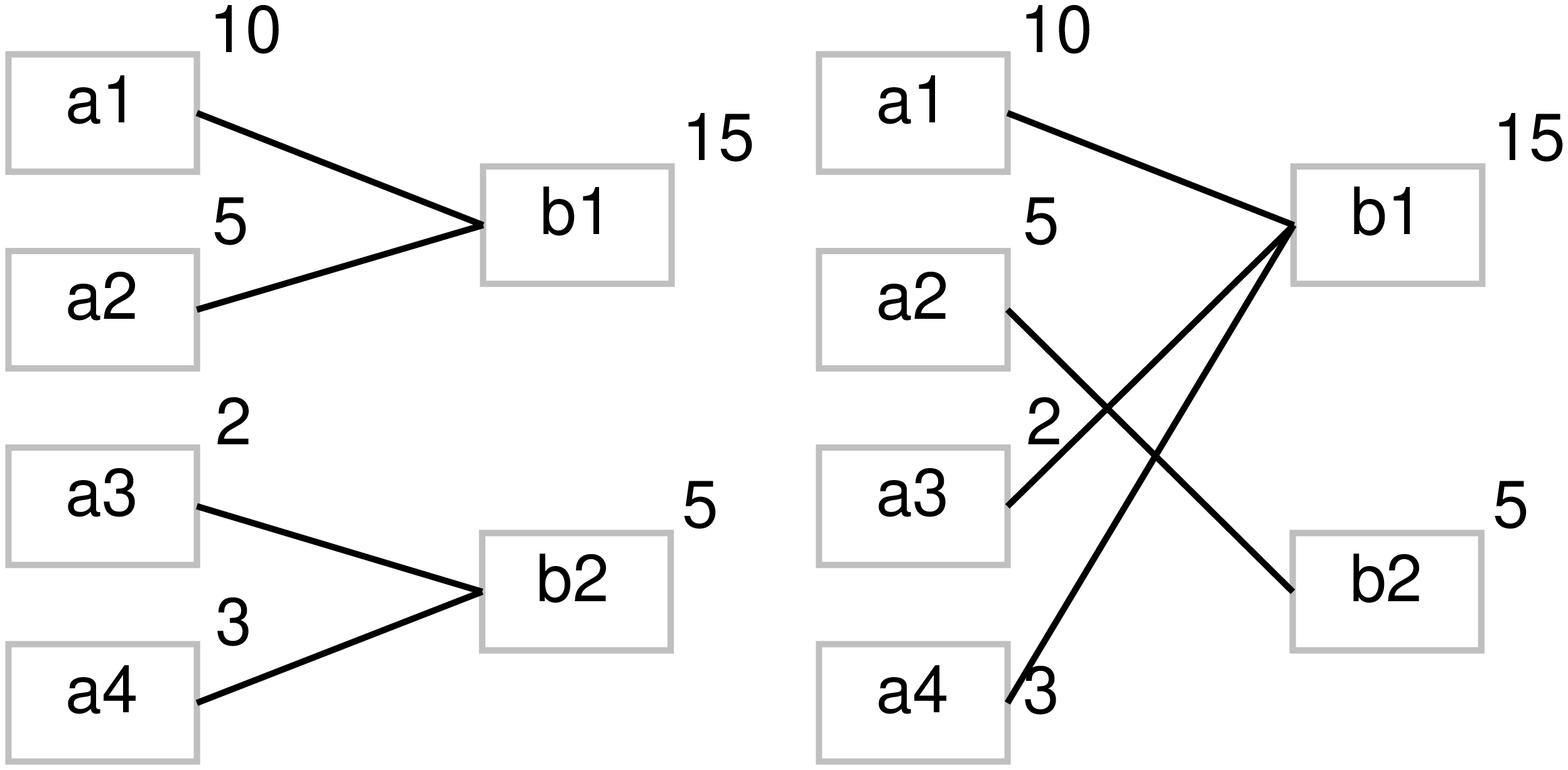}
%&
%\includegraphics[width=0.15\textwidth]{./figs/plaintextAttack2b.eps}
\\
(c) Freq+FD attack on $D_1$
& (d) Freq+FD attack on $D_2$
%&
%\\
%on $D_1$  
%&
%on $D_2$
%&  
\end{tabular}
\caption{\label{fig:pattack} Examples of the Freq+FD attack}
\end{small}
\end{center}
\vspace{-0.3in}
%\end{figure*}
\end{figure}

Without the access to the outsourced (and encrypted) data, the attacker may try to reconstruct (parts of) the original dataset $D$ based on the knowledge of the frequency of plaintext values and the $FDs$. In particular, the attacker knows that if there exists an FD $A\rightarrow B$, then there are two facts:
\begin{enumerate}
\item The number of unique instances of $A$ is no less than that of $B$; and 
\item For each value $a$ and $b$ of the attributes $A$ and $B$, it must be true that $f_a \leq f_b$. 
\end{enumerate}
Therefore, given a set of plaintext values $\cal A$ $(a_1, \dots, a_k)$ of attribute $A$, and a set of plaintext values $\cal B$ $(b_1, \dots, b_\ell)$ of attribute $B$ ($\ell\leq k$), the attacker looks for all possible assignments between $\cal A$ and $\cal B$ such that in each assignment: (1) each value $a_i\in\cal A$ is assigned a unique value $b_j\in\cal B$ such that $f_{a_i}\leq f_{b_j}$, and (2) for each each value $b_j\in\cal B$, it is assigned a set of values $a_1, \dots, a_t\in\cal A$ such that $\sum_{i=1}^{t}f_{a_i} = f_{b_j}$. It is possible that $t=1$, i.e., $b_j$ is associated with a single value $a_i$, where $f_{a_i}=f_{b_j}$. 
If a unique assignment is found, the attacker will reconstruct the plaintext value pairs $(a_1, b_i), \dots, (a_t, b_i)$ pairs, for each value $b_i\in\cal B$, with 100\% certainty. Example \ref{exp:1} gives more details of the {\bf Freq+FD} attack.

\vspace{-0.05in}
\begin{example}
\label{exp:1}
{\em 
Given a plaintext dataset $D_1$, assume that the attacker knows $D_1$ has two attributes $A$ and $B$, among which there exists a functional dependency $F: A\rightarrow B$. The attacker knows the data domain and frequency information of $D_1$ from its prior knowledge (shown in Figure \ref{fig:pattack} (a) in the format of ({\em value: frequency})). The attacker can get one single assignment (Figure \ref{fig:pattack} (b)) by launching the {\bf Freq+FD} attack. Following the assignment it can construct the following data pairs: $(a_1, b_1)$ of frequency 10, $(a_2, b_1)$ of frequency 6, $(a_3, b_2)$ of frequency 2, and $(a_4, b_2)$ of frequency 3 with 100\% certainty. 
Now consider another plaintext dataset $D_2$, which has the same functional dependency $F: A\rightarrow B$ but different frequency only on $a_2$ and $b_2$. The frequency information of $D_2$ is shown in Figure \ref{fig:pattack} (b). Now the attacker can get two assignments (Figure \ref{fig:pattack} (d)) by launching the {\bf Freq+FD} attack. In both assignments, the data pairs $(a_1, b_1)$ of frequency 10 always exist. Therefore the attacker has 100\% certainty that there are 10 data pairs $(a_1, b_1)$ in $D$. But it only can decide the existence of $(a_2, b_1)$ of frequency 5 with 50\% belief probability, similarly for $(a_3, b_2)$ of frequency 2, and $(a_4, b_2)$ of frequency 3. 
}\qed
\end{example}
\vspace{-0.05in}
In general, given a set of integers \{$f_1, \dots, f_k$\} and an integer $s$, finding whether there exists a subset whose sum is equal to $s$ is equivalent to the well-known subset sum problem. The {\em Freq+FD} attack is indeed a problem that is harder than the subset sum problem, as it aims to find $\ell$ non-overlapping subsets, each with sum equal to an integer $s_i (1\leq i\leq\ell)$. We have the following theorem to show the sufficient condition of the {\bf Freq+FD} attack that can succeed with 100\% certainty. 
\begin{theorem}
{\em 
Given a FD $A\rightarrow B$, for any data value $a_i$ of attribute $A$ and $b_j$ of attribute $B$ such that: (1)  $f_{a_i} = f_{b_j}$, and (2) both $f_{a_i}$ and $f_{b_j}$ are unique (i.e., the frequency of other instances are not $f_{a_i}$ and $f_{b_j}$), then the pair $(a_i, b_j)$ with frequency $f_{a_j}$ can be reconstructed by the {\bf Freq+FD} attack with 100\% certainty.}
\end{theorem}

To measure the feasibility of the {\bf Freq+FD} attack, we created 40 datasets with two columns $\cal{A}$ and $\cal{B}$, on which the $FD$ $\cal{A}$ $\rightarrow \cal{B}$ holds. The domain size of column $\cal{A}$ varies from 100 to 1000. The number of distinct values in $\cal{A}$ varies from 10 to 20. We design these datasets in two ways: (1) the frequency distribution is skewed; and (2) the frequency distribution is close to uniform. We implement and simulate the {\bf Freq+FD} attack on these datasets, and measure how likely the {\bf Freq+FD} attack can be successful. We define the successful attack rate as $p=\frac{x}{|D|}$, where $x$ is the number of records that can be reconstructed by the {\bf Freq+FD}  attack with 100\% certainty, and $|D|$ is the number of records in the original dataset.
%We show the successful attack rate of each configuration in Figure \ref{fig:attack}. 
We observe that the successful attack rate varies from 0.04 to 1. In particular, the skewed frequency distribution is more likely to face with the higher successful attack rate, as the number of valid mappings between the two attributes that satisfy the frequency constraint is smaller. We also observe that the smaller domain leads to higher successful rate as it has smaller number of mappings. 

We must note that our $F^2$ algorithm cannot provide any security guarantee against the {\bf Freq+FD} attack, as both the frequency knowledge and the $FDs$ are available to the attacker, no matter how $D$ is encrypted (as long as the encryption is FD-preserving). One possibility to defend against the  {\bf Freq+FD} attack is to use non FD-preserving encryption schemes, so that the attacker cannot discover FDs from the encrypted dataset. However this will disable schema refinement service at the outsourced dataset. 
}

\vspace{-0.05in}
\subsection{Without Kerckhoffs's Principle}
\label{sc:semi}

\nop{
We assume the attacker knows the set of plaintext values $\cal P$ and their exact frequency in $D$. Meanwhile he collects  the ciphertext values $\cal E$ and their exact frequency from the received dataset $\hat{D}$. Based on the collected  information the attacker constructs  a surjection from $\cal E$ to $\cal P$ such that each ciphertext value $e\in\cal E$ is mapped to a plaintext value $p\in P$ if they have the same frequency. 
First, let us consider the ciphertext values that map to real plaintext values in $D$. According to $F^2$, for each ciphertext value $e$, there always exists a set of ciphertext values that are of the same frequency. Let the number of such ciphertext values be $\ell$. These $\ell$ ciphertext values correspond to $k$ unique plaintext values in the same $ECG$. Therefore, the probability of associating any ciphertext value with a plaintext value based on their frequency is at most $\frac{1}{k}$. As we always require that $k\geq \frac{1}{\alpha}$, therefore, the probability $Prob(e \rightarrow p) \leq \alpha$. 
Next, we discuss the probability of breaking the ciphertext values that map to artificial plaintext values. 
According to the Step 4 that constructs artificial records (Section \ref{sc:fdfp}), each such ciphertext value is of the same frequency. There are at least $k = \lceil\frac{1}{\alpha}\rceil$ such unique ciphertext values. Thus, the probability of breaking any of such $k$ ciphertext values based on their frequency is at most $\frac{1}{k} = \alpha $.
 Therefore, {\em $F^2$} guarantees $\alpha$-security against the {\bf Freq+$\hat{D}$} attack.
 }

%To show that $F^2$ provides $\alpha$-security against the frequency analysis attack, we need to prove that it restricts the success probability for the adversary with frequency knowledge to play the experiment $Exp_{\mathcal{A}, \Pi}^{freq}$  in Section \ref{sc:attack} .
For any $e\in \mathcal{E}$ be a ciphertext value, let $G(e)=\{p| p\in \mathcal{P}, freq_{\mathcal{P}}(p)=freq_{\mathcal{E}}(e)\}$ be the set of distinct plaintext values having the same frequency as $e$. It has shown \cite{sanamrad2014randomly} that for any adversary $\mathcal{A}^{freq}$ and any ciphertext value $e$, the chance that the adversary succeeds the frequency analysis attack is 
$Pr[Exp_{\mathcal{A}, \Pi}^{freq}=1] = \frac{1}{|G(e)|}$, where $|G(e)|$ is the size of $G(e)$.
In other words, the size of $G(e)$ determines the success probability of $Exp_{\mathcal{A}, \Pi}^{freq}$.

Apparently, the scaling step of $F^2$ ensures that all the equivalence classes in the same $ECG$ have the same frequency. 
%Hence, for any encrypted equivalence class $EC'$, $G(EC')=\{EC| EC in the same ECG as EC'\}$.
Hence, for any encrypted equivalence class $EC'$, there are at least $|ECG|$ plaintext $EC$s having the same frequency.
Recall that the way we form the equivalence class groups does not allow any two equivalence classes in the same $ECG$ to have the same value on any attribute. 
Therefore, for any attribute $A$, a $ECG$ contains $k$ distinct plaintext values on $A$, where $k$ is the size of $ECG$. 
Thus for any $e\in \mathcal{E}$, it is guaranteed that $|G(e)| = k$. 
As $k\geq [\frac{1}{\alpha}]$, it is guaranteed that $G(e)\geq [\frac{1}{\alpha}]$.
In this way, we have $Pr[Exp_{\mathcal{A}, \Pi}^{freq}=1]\leq \alpha$. Thus $F^2$ is $\alpha$-secure against the frequency analysis attack.

\nop{
\begin{figure*}[ht]
\begin{center}
\begin{small}
\begin{tabular}{ccc}
\begin{tabular}{| c | c | c | c |}\hline
\hline
A & $freq $ & B & $freq$\\
\hline
$a_1$ & $f_1$ & $b_1$ & $f_3+f_4$\\\hline
$a_2$ & $f_2$ & $b_2$ & $f_1$\\\hline
$a_3$ & $f_3$ & $b_3$ & $f_2$\\\hline
$a_4$ & $f_4$ & $ $ & $ $\\\hline
\end{tabular}
&
\begin{tabular}{| c | c | c | c |}\hline
\hline
$Enc(A)$ & $Enc(B)$ & $freq$\\
\hline
$\hat{a}_1^1$ & $\hat{b}_2^1$ & $f^1$\\\hline
$\hat{a}_3^1$ & $\hat{b}_1^1$ & $f^1$\\\hline
$\hat{a}_3^2$ & $\hat{b}_1^2$ & $f^1$\\\hline
\end{tabular}
&
\begin{tabular}{| c | c | c | c |}\hline
\hline
$Enc(A)$ & $Enc(B)$ & $freq$\\
\hline
$\alpha_2^1$ & $\hat{b}_3^1$ & $f^2$\\\hline
$\alpha_4^1$ & $\hat{b}_1^3$ & $f^2$\\\hline
$\alpha_4^2$ & $\hat{b}_1^4$ & $f^2$\\\hline
\end{tabular}
\\
(a) Frequency knowledge $D$ ($FD: A\rightarrow B$)& (b) $Enc(G_1)$: the ciphertext values & (e) $Enc(G_2)$: the ciphertext values\\
& and frequency of Group 1& and frequency of Group 2
\end{tabular}
\caption{\label{fig:attack} An example of possible attacks.}
\end{small}
\end{center}
\vspace{-0.3in}
%\end{figure*}
\end{figure*}

However, the above reasoning holds only in a single $ECG$. If there is some relationship between the ciphertext values in different $ECGs$, the attacker may utilize the information to infer more data. Typically, if $e_p = e_q$ for $e_p \in Enc(ECG_i)$ and $e_q \in Enc(ECG_j)$ with $i \neq j$, the attacker would know that $e_p$ and $e_q$ correspond to one plaintext value $v$, and $v$ exists in different $ECGs$. For example, in Figure \ref{fig:attack}(b) and (e), if $\hat{b}_1^1=\hat{b}_1^3$, the attacker knows that they have the same plaintext value $b$, and $b$ corresponds to two different values in attribute $A$. With the frequency knowledge in Figure \ref{fig:attack}(a), it's easy for the attacker to infer that $\hat{b}_1^1$ and $\hat{b}_1^3$'s plaintext value is $b_1$. So we should forbid the attacker from linking $ECG$s by insisting that the same plaintext value in different $ECG$s should have different ciphertext values. 
}
\vspace{-0.05in}
\subsection{Under Kerckhoffs's principle}
\label{sc:mali}
We assume the attacker knows the details of the $F^2$ algorithm besides the frequency knowledge. We discuss how the attacker can utilize such knowledge to break the encryption.
We assume that the attacker does not know the $\alpha$ and $\varpi$ values that data owner uses in $F^2$.  
Then the attacker can launch the following 4-step procedure.

\noindent{\bf Step 1: Estimate the split factor $\varpi$.} The attacker finds the maximum frequency $f_{m}^P$ of plaintext values and the maximum frequency $f_{m}^E$ of ciphertext values. Then it calculates $\varpi' = \frac{f_{m}^E}{f_{m}^P}$. It is highly likely that $\varpi' = \varpi$. 

\noindent{\bf Step 2: Find $ECGs$.} The attacker applies Step 2.1 of $F^2$ bucketizes $\cal E$ by grouping ciphertext values of the same frequency into the same bucket. Each bucket corresponds to one $ECG$. 

\noindent{\bf Step 3: Find mappings between $ECGs$ and plaintext values.} The attacker is aware of the fact that for a given ciphertext value $e$, its frequency $f_{\hat{D}}(e)$ must satisfy that $f_{\hat{D}}(e)\geq \varpi f_{D}(p)$, where $p$ is the corresponding plaintext value of $e$. Following this reasoning, for any $ECG$ (in which all ciphertext values are of the same frequency $f$), the attacker finds all plaintext values $\cal P'$ such that $\forall P\in\cal P'$,  $\varpi f_{D}(p)\leq f$. 

\noindent{\bf Step 4: Find mappings between plaintext and ciphertext values.} For any $ECG$, the attacker maps any ciphertext value in $ECG$ to a plaintext value in the candidate set returned by Step 3. Note the attacker can run $F^2$ to find the optimal split point of the given $ECG$. 

Next, we analyze the probability that the attacker can map a ciphertext value $e$ to a plaintext value $p$ by the 4-step procedure above. It is possible that the attacker can find the correct mappings between $ECGs$ and their plaintext values with 100\% certainty by Step 1 - 3. Therefore, we mainly analyze the probability of Step 4. Given an  $ECG$ that matches to $k$ plaintext values, let $y$ be the number of its unique ciphertext values. Then the number of possible mapping of $y$ ciphertext values (of the same frequency) to $k$ plaintext values is $\binom{y}{k}k^{y-k}$. Out of these mappings, there are $\binom{y-1}{k-1}k^{y-k-1}$ mappings that correspond to the mapping $e\rightarrow p$. Therefore, 
\vspace{-.05in}
\[Prob(e\rightarrow p)=\frac{\binom{y-1}{k-1}k^{y-k-1}}{\binom{y}{k}k^{y-k}} = \frac{1}{y}.\]
Assume in the given $ECG$, $k'\leq k$ plaintext values are split by $F^2$. The total number of ciphertext values of the given $ECG$ is $y = \varpi k' + k - k'$. It is easy to compute that $y\geq k$. As we always guarantee that $k\geq\frac{1}{\alpha}$, 
$Prob(e\rightarrow p) = \frac{1}{y} \leq \frac{1}{k} \leq \alpha.$
Therefore, $F^2$ guarantees $\alpha$-security against the frequency analysis attack under Kerckhoffs's principle. 

\nop{
We are aware that the attacker is able to distinguish the artificial records by Step 4 (Section \ref{sc:fdfp}) as these records always consist of values of frequency one. However, we argue that identifying the artificial records does not help to break the encryption on real records, since: (1) these artificial records do not share any value of the real records, and (2) these records are constructed for eliminating false positive FDs only. Thus removing the artificial records from the outsourced dataset cannot improve $Prob(e\rightarrow p)$ for any ciphertext value $e$ and plaintext value $p$.
}

 %So he could initiate the known-scheme attack. The attacker knows our split factor is decided by $\varpi$. He is able to "guess" the approximate value of $\varpi$. Let $f_{max}$ denote $max\{freq(v)|v \in dom(A), A \in LHS(F)\}$ and $f_{max}'$ be $max\{freq(e) | c \in Enc(ECG)\}$. The attacker could guess $\overline{\varpi}=\frac{f_{max}}{f_{max}'}$. This is because according to our split scheme, we have a good probability to split the $ECG$ containing a value with the largest frequency and $f_{max}' \geq \frac{1}{\varpi}*f_{max}$. If the attacker gets a close guess of $\varpi$ and find the frequency of an $Enc(ECG)=\{e_1, \dots, e_m\}$ equal to $[\frac{1}{\overline{\varpi}}*f_{max}]$, he knows that the plaintext value $v$ with frequency $f_{max}$ exists in $ECG$ and it is splitted. The probability to successfully map a ciphertext value in $Enc(ECG)$ to $v$ is $Prob(e_i \rightarrow v) = \frac{\varpi}{m}$. The worst case happens when we only split $v$ in this $ECG$. Following our example in Figure \ref{fig:attack}, if we split both $G_1$ and $G_2$, $f^1=[\frac{f_3}{\varpi}]$, $f^2=\frac{f_4}{\varpi}$. $f_{max}'=f^2$. By $\overline{\varpi}=\frac{f_{max}}{f_{max}'}$, the attacker gets the exact value of $\varpi$. Because $f^1=\frac{f_3}{\overline{\varpi}}$ and $f^2=\frac{f_4}{\overline{\varpi}}$, the attacker knows that $a_3$ exists in $G_1$ and $a_4$ exists in $G_2$. So $Prob(e_i \rightarrow v) \leq \frac{\varpi}{\varpi+k-1}$. As we require $k\geq [\frac{(1-\alpha)*{\varpi}}{\alpha}]+1$, it's always true that $Prob(e_i \rightarrow v) \leq \alpha$.

\vspace{-0.1in}
\section{Experiments}
\label{sc:exp}
%We ran an extensive set of experiments to evaluate the efficiency of our approach. 
In this section, we discuss our experiment results and provide the analysis of our observations.

\subsection{Setup}
\noindent {\bf Computer environment.} We implement our algorithm in Java. All the experiments are executed on a PC with 2.5GHz i7 CPU and 60GB memory running Linux.

\noindent {\bf Datasets.} We execute our algorithm on two TPC-H benchmark datasets, namely the {\em Orders} and {\em Customer} datasets, and one synthetic dataset. More details of these three datasets can be found in Table \ref{tb:data}. 
%The {\em Orders} dataset contains 9 attributes and 1.5 million records (size 1.67GB). 
{\em Orders} dataset contains nine maximal attribute sets $MASs$. 
All $MASs$ overlap pairwise. Each $MAS$ contains either four or five attributes. 
%The {\em Customer} dataset includes 21 attributes and 960K records (size 282MB). 
There are fifteen $MAS$s in {\em Customer} dataset. The size of these $MASs$ (i.e., the number of attributes) ranges from nine to twelve. All $MAS$s overlap pairwise. 
%The synthetic dataset contains 7 attributes and 4 million records (size 224MB). 
The synthetic dataset has two $MAS$s, one of three attributes, while the other of six attributes. The two $MASs$ overlap at one attribute. 

\begin{table}[h!]
\centering
\begin{tabular}{|c|c|c|c|}
\hline
Dataset & \# of attributes & \# of tuples & size\\
        &        &   (Million)            & \\\hline
Orders & 9 & 15 & 1.64GB\\\hline
Customer & 21 & 0.96 & 282MB\\\hline
Synthetic & 7 & 4 & 224MB\\\hline
\end{tabular}
\caption{Dataset description}
\label{tb:data}
\end{table}
%\vspace{-0.1in}
%\noindent {\bf Parameters.} We evaluate our approach against various $\alpha$ values for $\alpha$-security, and various data sizes. We construct a set of datasets from the {\em Orders} dataset of size $0.3M$, $0.6M$, $0.9M$ and $1.2M$. These datasets are generated by picking a certain number of records from the original {\em Orders} dataset. We build a set of datasets from the {\em Customer} and synthetic datasets that contain $60K$, $120K$, $240K$, $480K$, $960K$ and $512K$, $1024K$, $2048K$, $4096K$ records respectively. 
%To eliminate duplicate tuples, we append a token, which is unique for each copy, to all values in each record. In this way, the $MAS$s in these datasets are the same as those in the original dataset. So we are able to evaluate the scalability of our encryption scheme regarding the number of tuples, with the same $MAS$s.

\nop{
\begin{table}[h!]
\centering
\begin{tabular}{|c|c|c|c|}
\hline
Dataset & $m$ & $n$ (Million) & size (MB) \\\hline
\multirow{ 5}{*}{Orders} & \multirow{ 5}{*}{9} & $3$ & $333$\\
& & $6$ & $669$ \\
& & $9$ & $1005$ \\
& & $12$ & $1340$ \\
& & $15$ & $1676$ \\
\hline
\multirow{ 4}{*}{Synthetic} & \multirow{ 5}{*}{7} & $0.5$ & $25$\\
& & $1$ & $53$ \\
& & $2$ & $110$ \\
& & $4$ & $224$ \\
\hline
\multirow{ 4}{*}{Customer} & \multirow{ 5}{*}{21} & $0.12$ & $33$ \\ 
& & $0.24$ & $68$ \\
& & $0.48$ & $139$ \\
& & $0.96$ & $282$ \\
\hline
\end{tabular}
\caption{Data description}
\label{tb:data}
\end{table}
}
\noindent{\bf Evaluation.} We evaluate the efficiency and practicality of our encryption scheme according to the following
criteria:
\vspace{-0.05in}
\begin{ditemize}
\item Encryption time: the time for the data owner to encrypt the dataset (Sec. \ref{sc:etime});
\item Space overhead: the amounts of artificial records added by $F^2$ (Sec. \ref{sc:arti}); 
\item Outsourcing versus local computations: (1) the time of discovering FDs versus encryption by $F^2$, and (2) the FD discovery time on the original data versus that on the encrypted data (Sec. \ref{sc:local}). 
\end{ditemize}
\vspace{-0.05in}
\noindent{\bf Baseline approaches.} We implement two baseline encryption methods ({\em AES} and {\em Paillier}) to encode the data at cell level. The {\em AES} baseline approach uses the well-known AES algorithm for the deterministic encryption. We use the implementation of AES in the {\em javax.crypto} package. The {\em Paillier} baseline approach is to use the asymmetric Paillier encryption for the probabilistic encryption. We use the {\em UTD Paillier Threshold Encryption Toolbox}\footnote{\url{http://cs.utdallas.edu/dspl/cgi-bin/pailliertoolbox/}.}. Our probabilistic approach is implemented by combining a random string (as discussed in Section \ref{sc:sd}) and the AES algorithm. We will compare the time performance of both {\em AES} and {\em Paillier} with $F^2$. 

\subsection{Encryption Time}
\label{sc:etime}
\begin{figure}[ht]
  \begin{center}
    \begin{tabular}{cc}
      \includegraphics[width=0.23\textwidth]{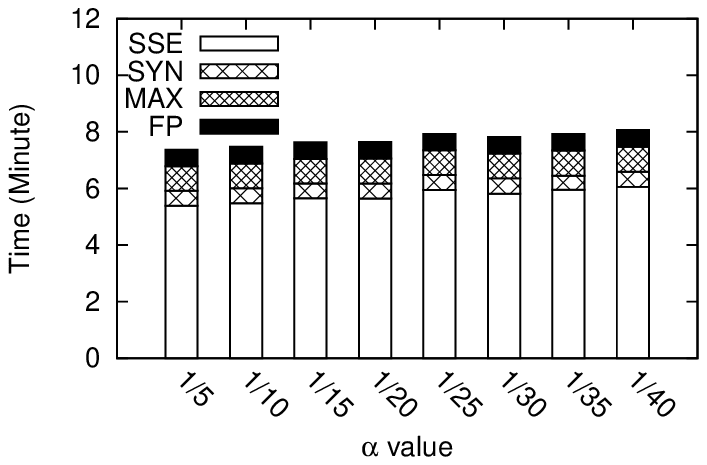}
      &
      \includegraphics[width=0.23\textwidth]{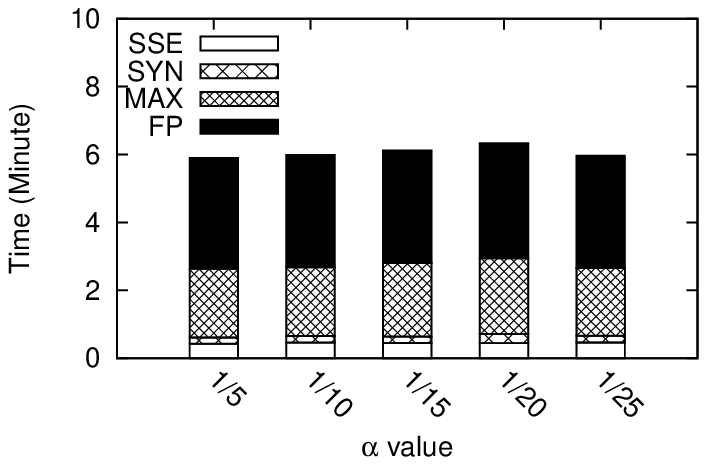}
      \\
      (a) Synthetic dataset (53MB)
      &
      (b) Orders dataset (0.325GB)
    \end{tabular}
    \vspace{-0.15in}
    \caption{\small \label{fig:time_vs_k} Time performance for various $\alpha$}
    \vspace{-0.2in}
  \end{center}
\end{figure}

In this section, we measure the time performance of $F^2$ to encrypt the dataset. First, we evaluate the impact of security threshold $\alpha$ on the running time. We measure the time of our four steps of the algorithm: (1) finding maximal attribute sets (MAX), (2) splitting-and-scaling encryption (SSE), (3) conflict resolution (SYN), and (4) eliminating false positive FDs (FP), individually. We show our results on both {\em Orders} and the synthetic datasets in Figure \ref{fig:time_vs_k}. 
First, we observe that for both datasets, the time performance does not change much with the decrease of $\alpha$ value. This is because the running time of the MAX, SYN and FP steps is independent on $\alpha$.
In particular, the time of finding $MASs$ stays stable with the change of $\alpha$ values, as its complexity relies on the data size, not $\alpha$. Similarly, the time performance of FP step is stable with various $\alpha$ values, as its complexity is only dependent on the data size and data schema. The time performance of SYN step does not vary with $\alpha$ value because we only need to synchronize the encryption on the original records, without the worry about the injected records. It is worth noting that on the {\em Orders} dataset, the time took by the SYN step is negligible. This is because the SYN step leads to only $24$ artificial records on the {\em Orders} dataset (of size $0.325$GB). 
Second, the time of SSE step grows for both datasets when $\alpha$ decreases (i.e., tighter security guarantee). This is because smaller $\alpha$ value requires larger number of artificial equivalence classes to form $ECG$s of the desired size. This addresses the trade-off between security and time performance. 
We also observe that the increase in time performance is insignificant. For example, even for {\em Orders} dataset of size $0.325$GB, when $\alpha$ is decreased from $0.2$ to $0.04$, the execution time of the SSE step only increases by $2.5$ seconds. This shows that $F^2$ enables to achieve higher security with small additional time cost on large datasets. 
We also observe that different steps dominate the time performance on different datasets: the SSE step takes most of the time on the synthetic dataset, while the MAX and FP steps take the most time on the {\em Orders} dataset.  This is because the average number of $ECs$ of all the $MAS$s in the synthetic dataset is much larger than that of the {\em Orders} dataset (128,512 v.s. 1003). Due to the quadratic complexity of the SSE step with regard to the number of $ECs$, the SSE step consumes most of the time on the synthetic dataset. 

\nop{
Second, we observe that the dominant time factor on {\em Adult} dataset is the time of FP step, while on the synthetic dataset is the time of SSE step. This is because the {\em Adult} dataset has 10 $MAS$s that contain more than 10 attributes. Hence, we need to go through the large FD lattice to eliminate false positive FDs. On the contrary, the synthetic dataset only contains 2 $MAS$s of size 3 and 6. Therefore, eliminating false positive FDs from the synthetic dataset is much faster than that of the Adult dataset. 
}
  \vspace{-0.1in}

\begin{figure}[ht]
  \begin{center}
    \begin{tabular}{cc}
      \includegraphics[width=0.23\textwidth]{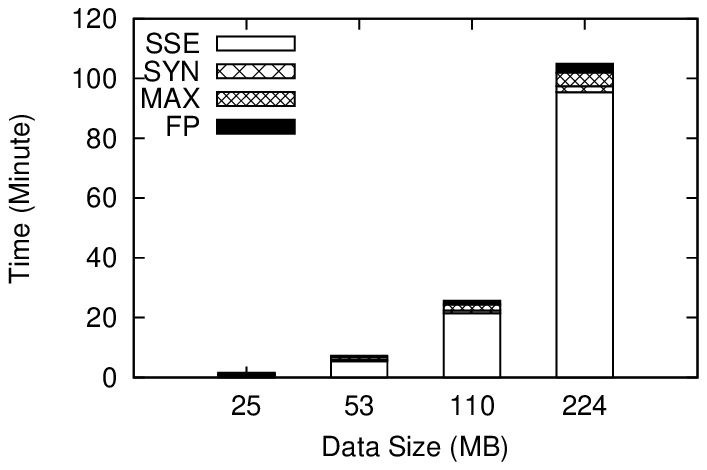}
      &
      \includegraphics[width=0.23\textwidth]{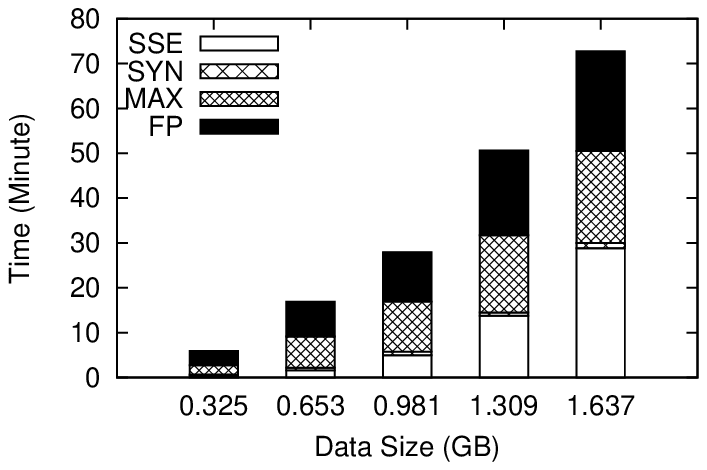}
      \\
      (a) Synthetic dataset ($\alpha$=0.25)
      &
      (b) Orders dataset ($\alpha$=0.2)
    \end{tabular}
    \vspace{-0.15in}
    \caption{\small \label{fig:time_vs_size} Time performance for various data sizes}
    \vspace{-0.2in}
  \end{center}
\end{figure}

Second, to analyze the scalability of our approach, we measure the time performance for various data sizes. The results are shown in Figure \ref{fig:time_vs_size}. It is not surprising that the time performance of all the four steps increases with the data size. 
We also notice that, on both datasets, the time performance of the SSE step is not linear to the data size. 
This is due to the fact that the time complexity of the SSE step is quadratic to the number of $EC$s. With the increase of the data size, the average number of $EC$s increases linearly on the synthetic dataset and super-linearly on the {\em Orders} dataset. Thus we observe the non-linear relationship between time performance of SSE step and data size.
On the synthetic dataset, the dominant time factor is always the time of the SSE step. This is because of the large average number of $EC$s (can be as large as 1 million) and the quadratic complexity of the SSE step. In contrast, the average number of $EC$s is at most $11,828$ on the {\em Orders} dataset. So even though the synthetic dataset has fewer and smaller $MAS$s than the {\em Orders} dataset, the vast difference in the number of $EC$s makes the SSE step takes the majority of the time performance on the synthetic dataset.

To sum up the observations above, our $F^2$ approach can be applied on the large datasets with high security guarantee, especially for the datasets that have few number of $EC$s on the $MAS$s. For instance, it takes around 30 minutes for $F^2$ to encrypt the {\em Orders} dataset of size 1GB with security guarantee of $\alpha=0.2$. 

%\vspace{-0.1in}

%\vspace{-0.2in}
\begin{figure}[ht]
  \begin{center}
    \begin{tabular}{cc}
      \includegraphics[width=0.23\textwidth]{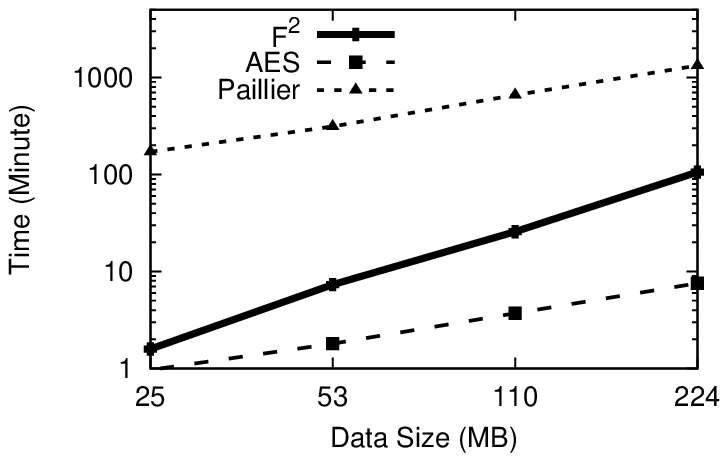}
      &
      \includegraphics[width=0.23\textwidth]{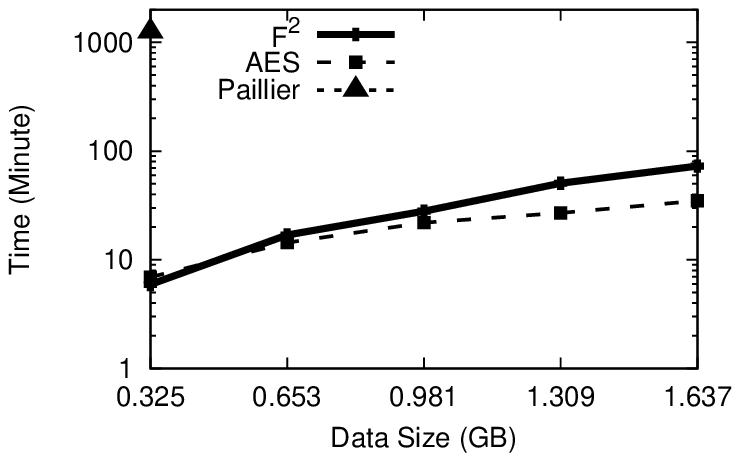}
      \\
      (a) Synthetic dataset ($\alpha$=0.25)
      &
      (b) Orders dataset ($\alpha$=0.2)
    \end{tabular}
    \vspace{-0.15in}
    \caption{\small \label{fig:time_comparison} Time performance Comparison}
    \vspace{-0.2in}
  \end{center}
\end{figure}

We also compare the time performance of $F^2$ with the two baseline methods. 
%In this set of experiments, we take into account the time that $F^2$ spends on doing encryption. 
The result is shown in Figure \ref{fig:time_comparison}. 
It is not surprising that $F^2$ is slower than {\em AES}, as it has to handle with the FD-preserving requirement. 
On the other hand, even though $F^2$ has to take additional efforts to be FD-preserving (e.g., finding $MASs$, splitting and scaling, etc.), its time performance is much better than {\em Paillier}. This shows the efficiency of our probabilistic encryption scheme. 
It is worth noting that on the {\em Orders} dataset, {\em Paillier} takes 1247.27 minutes for the data of size 0.325GB, and cannot finish within one day when the data size reaches 0.653GB. Thus we only show the time performance of {\em Paillier} for the data size that is below 0.653GB. 

\begin{figure}[ht]
  \begin{center}
    \begin{tabular}{cc}
      \includegraphics[width=0.23\textwidth]{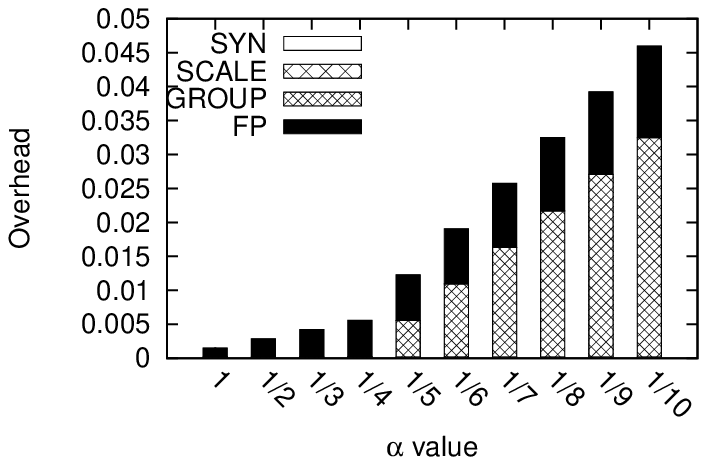}
      &
      \includegraphics[width=0.23\textwidth]{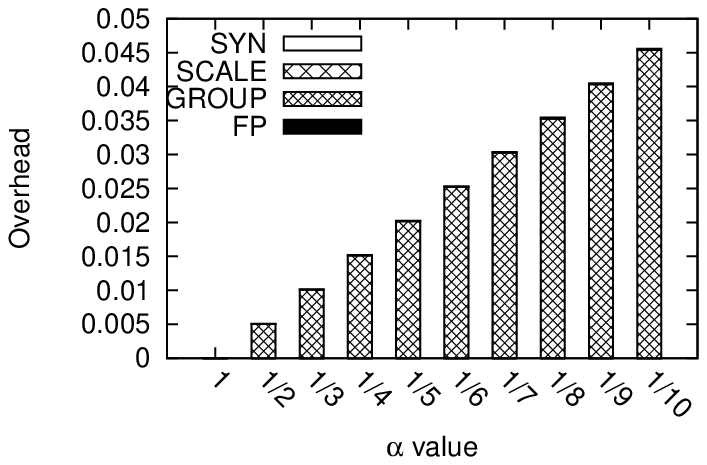}
      \\
      (a) Various $\alpha$ values
      &
      (b) Various $\alpha$ values
      \\
       ({\em Customer} 73MB)
       &
       ({\em Orders} 0.325GB)
      \\
      \includegraphics[width=0.23\textwidth]{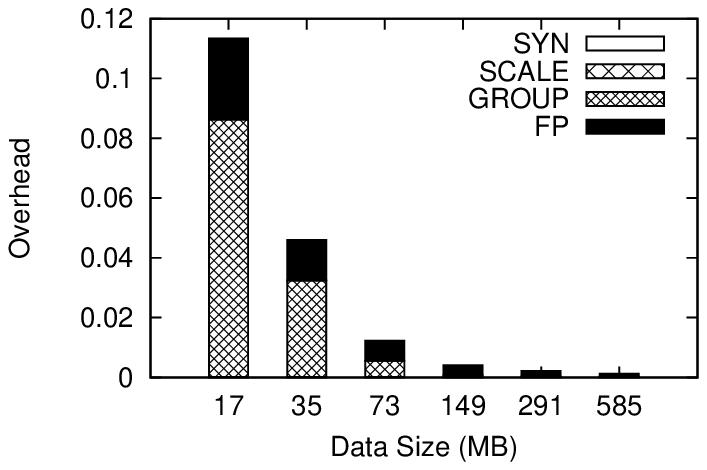}
      &
      \includegraphics[width=0.23\textwidth]{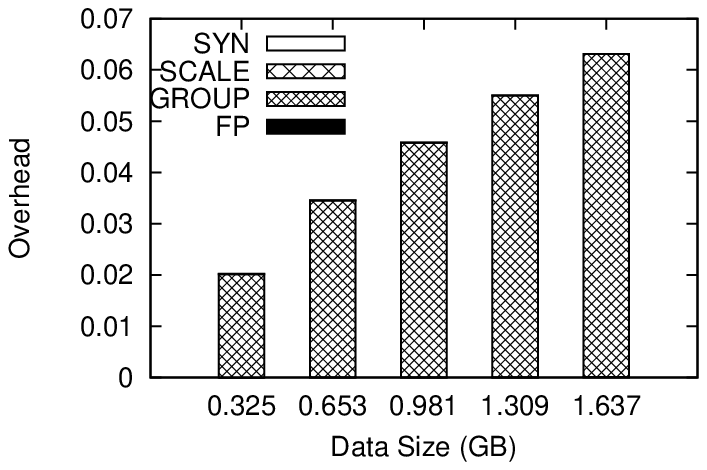}
      
      \\
      (c) Various data size 
      &
      (d) Various data size
      \\
      ({\em Customer} $\alpha=0.2$)
      &
      ({\em Orders} $\alpha$=0.2)
   \end{tabular}
    \vspace{-0.15in}
    \caption{\small \label{fig:ratio} Amounts of Artificial Records Added by $F^2$}
    \vspace{-0.3in}
  \end{center}  
\end{figure}

\subsection{Amounts of Artificial Records}
\label{sc:arti}
In this section, we measure the amounts of the artificial records added by $F^2$ on both the {\em Orders} and {\em Customer} datasets. We measure the amounts of the artificial records added by Step 2.1 grouping (GROUP), Step 2.2 splitting-and-scaling encryption (SCALE), Step 3 conflict resolution (SYN), and Step 4 eliminating false positive FDs (FP) individually. For each step, we measure the data size before and after the step, let them be $s$ and $s'$, and calculate the {\em space overhead} $r=\frac{s'-s}{s}$.

On the {\em Customer} dataset, we measure the overhead under various $\alpha$ values in Figure \ref{fig:ratio} (a). We observe that the GROUP and FP steps introduce most of the overhead. The overhead increases when $\alpha$ decreases, since smaller $\alpha$ value requires larger $ECGs$, and thus more collision-free equivalence classes. But in general, the overhead is very small, always within in 5\%. This is because the domain size of the attributes in $MAS$s of the {\em Customer} dataset is large. For instance, both the {\em C\_Last} and {\em C\_Balance} attribute have more than 4,000 unique values across 120,000 records. Under such setting, different $MASs$ are unlikely collide with each other. As a consequence, the number of artificial records added by the GROUP step is very small. 
When $\alpha < 0.2$, the GROUP step does not even introduce any overhead.
The overhead brought by the FP step increases with the decrease of $\alpha$ value. This is because for each maximum false positive FD, $F^2$ inserts $2k$ artificial records, where $k=\lceil\frac{1}{\alpha}\rceil$. As $k$ increases, the number of inserted records decreases. But even for small $\alpha$ value such as $\frac{1}{10}$, the overhead brought by the FP step is still very small (around 1.5\%). In any case, the space overhead of $F^2$ on the {\em Customers} dataset never exceeds 5\%. 
We also measure the space overhead with various $\alpha$ values on the {\em Orders} dataset. The result is shown in Figure \ref{fig:ratio} (b). First, the GROUP step adds dominant amounts of new records. The reason is that the domain size of attributes in $MAS$s on the {\em Orders} dataset is very small. For example, among the 1.5 million records, the {\em OrderStatus} and {\em OrderPriority} attributes only have 3 and 5 unique values respectively. Therefore the $ECs$ of these attributes have significant amounts of collision. This requires $F^2$ to insert quite a few artificial records to construct the $ECGs$ in which $ECs$ are collision-free. Nevertheless, the amounts of artificial records is negligible compared with the number of original records. The space overhead is 4.5\% at most. 
%We also observe that the overhead increases when $\alpha$ decreases as the price to pay higher security. The overhead brought by the other three steps is negligible due to the large overhead incurred by the GROUP step. 

In Figure \ref{fig:ratio} (c) and (d), we show the space overhead of both datasets of various sizes. On the {\em Customer} dataset, the overhead reduces with the increase of data size. Such overhead decreases for both GROUP and FP steps. Next we give the reasons. Regarding the GROUP step, this is because in the {\em Customer} dataset, the collision between $ECs$ is small. It is more likely that the GROUP step can find collision-free $ECs$ to form the $ECG$s, and thus smaller number of the injected artificial records. Regarding the FP step, its space overhead decreases when the data size grows is due to the fact that the number of artificial records inserted by the FP step is independent of the data size. This makes the number of artificial records constant when the data size grows. Therefore, the overhead ratio decreases for larger datasets. 
On the {\em Orders} dataset, again, the GROUP step contributes most of the injected artificial records. However, contrary to the {\em Orders} dataset, the overhead increases with the data size. This is because the $ECs$ of {\em Orders} have significant amounts of collision. Thus, the number of $ECs$ increases quadratically with the dataset size. Therefore, the space overhead of the GROUP step increases for larger datasets.

Combining the observations above, our $F^2$ method introduces insignificant amounts of artificial records. For instance, the amounts of artificial records takes at most 6\% for the {\em Orders} dataset, and 12\% for the {\em Customer} dataset. 

%\indent

\nop{
We also measure the encryption overhead for various data sizes. The result is shown in Figure \ref{fig:ratio_vs_size}. The two datasets show opposite patterns: for {\em Adult} dataset, the total overhead decreases, while for the synthetic dataset, the total overhead  increases for larger datasets. This is because for 
{\em Adult} dataset, Step 1 GROUP contributes dominant amounts of new records. Larger datasets lead to more equivalence classes, and thus relatively smaller number of new records to form the $ECG$. On the other hand, for the  
synthetic dataset, the new records mainly comes from Step 3 SYN. Larger datasets bring more conflicting equivalence classes, and thus more new records. 
%In Figure \ref{fig:ratio_vs_size}, we show the result on the {\em Adult} dataset. We observe that when the data size increases, the ratio decreases. This is because with more records, the number of added records in the grouping step becomes smaller. {\bf ???}
\vspace{-0.1in}
\begin{figure}[ht]
  \begin{center}
    \begin{tabular}{cc}
      \includegraphics[width=0.23\textwidth]{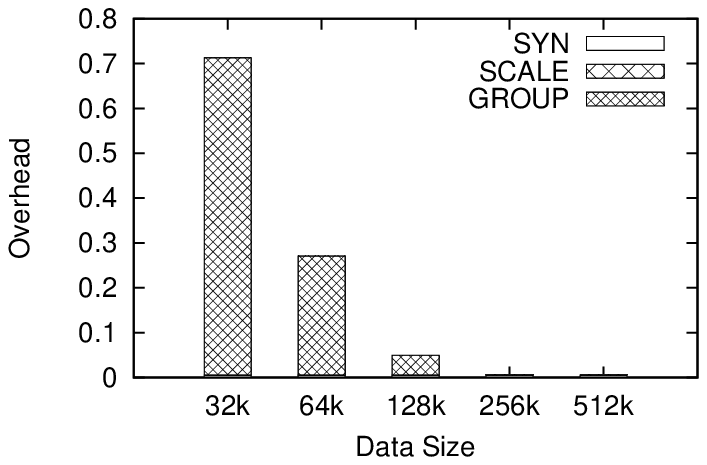}
      &
      \includegraphics[width=0.23\textwidth]{./exp/artificial_ratio_vs_size.eps}
      \\
      (a) Adult dataset ($\alpha$=0.2)
      &
      (b) Synthetic dataset ($\alpha$=0.25)
    \end{tabular}
    \vspace{-0.1in}
    \caption{\small \label{fig:ratio_vs_size} Encryption overhead for various data sizes}
    \vspace{-0.2in}
  \end{center}
\end{figure}
}
%\vspace{-0.1in}
\subsection{Outsourcing VS. Local Computations}
\label{sc:local}
First, we compare the data owner's performance of finding $FDs$ locally and encryption for outsourcing. We implemented the TANE algorithm \cite{huhtala1999tane} and applied it on our datasets. First, we compare the time performance of finding $FDs$ locally (i.e. applying TANE on the original dataset $D$) and outsourcing preparation (i.e., encrypting $D$ by $F^2$). It turns out that finding $FDs$ locally is significantly slower than applying $F^2$ on the synthetic dataset. For example, TANE takes 1,736 seconds on the synthetic dataset whose size is 25MB to discover FDs, while $F^2$ only takes 2 seconds. 

%\vspace{-0.1in}
\begin{figure}[ht]
  \begin{center}
    \begin{tabular}{cc}
      \includegraphics[width=0.23\textwidth]{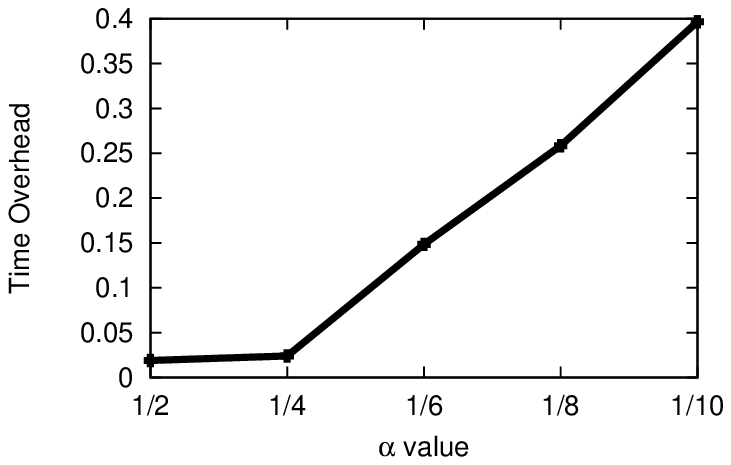}
      &
      \includegraphics[width=0.23\textwidth]{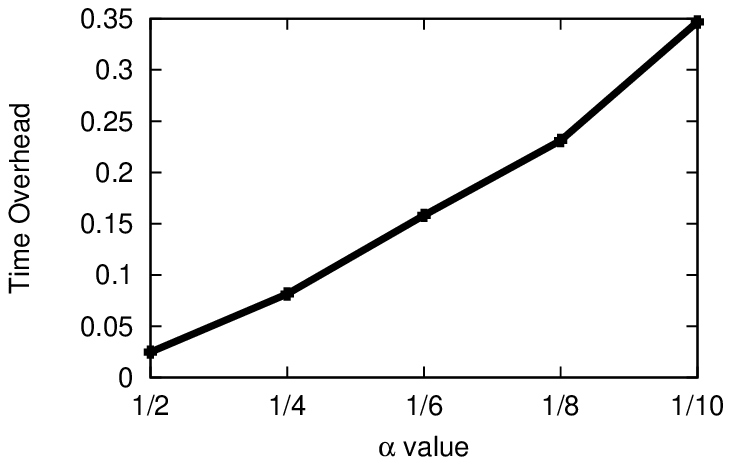}
      \\
      (a) Customer dataset (73MB)
			%(a) Customer dataset
      &
      (b) Orders dataset (0.325GB)
			%(b) Orders dataset 
    \end{tabular}
    \vspace{-0.15in}
    \caption{\small \label{fig:mas_vs_a} Dependency Discovery Time Overhead}
    \vspace{-0.2in}
  \end{center}
\end{figure}

Second, we compare the performance of discovering $FD$s from the original and encrypted data, for both {\em Customer} and {\em Orders} datasets. We define the {\em dependency discovery time overhead} $o=\frac{T'-T}{T}$, where $T$ and $T'$ are the time of discovering $FDs$ from $D$ and $\hat{D}$ respectively. The result is shown in Figure \ref{fig:mas_vs_a}. For both datasets,
the time overhead is small. It is at most 0.4 for the {\em Customers} dataset and 0.35 for the {\em Orders} dataset. Furthermore, 
 the discovery time overhead increases with the decrease of $\alpha$ value. This is because with smaller $\alpha$, the GROUP and FP steps insert more artificial records to form $ECG$s for higher security guarantee. Consequently the FD discovery time increases. This is the price to pay for higher security guarantee. 

\nop{
\begin{figure}[ht]
  \begin{center}
    \begin{tabular}{cc}
      \includegraphics[width=0.23\textwidth]{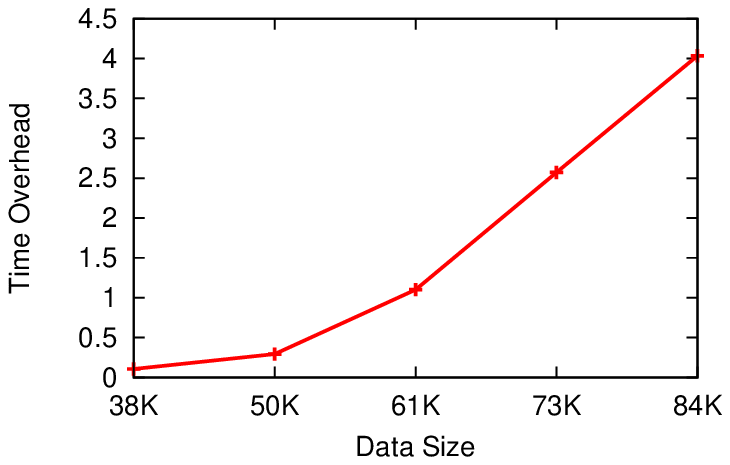}
      &
      \includegraphics[width=0.23\textwidth]{./exp/artificial_mas_vs_size.eps}
      \\
      (a) Adult dataset (32K)
      &
      (b) Synthetic dataset (64K)
    \end{tabular}
    \vspace{-0.1in}
    \caption{\small \label{fig:mas_vs_size} Time overhead for various data sizes}
    \vspace{-0.1in}
  \end{center}
\end{figure}
}

%\subsection{Simulation of Freq+FD Attack}
%We implement and simulate the {\bf Freq+FD} attack (Section \ref{sc:security}) on {\em Adult} dataset, and measure how likely the {\bf Freq+FD} attack can be successful against the real-world datasets. We define the successful attack rate as $p=\frac{x}{|D|}$, where $x$ is the number of records that can be reconstructed by the {\bf Freq+FD}  attack with 100\% certainty, and $|D|$ is the number of records in the original dataset. We create 6 datasets by picking two columns denoted as $\cal{A}$ and $\cal{B}$ from the {\em Adult} dataset whose instances are customized to satisfy the given functional dependency $\cal{A}$ $\rightarrow \cal{B}$. The size of the dataset varies from {\bf ???} to {\bf ???}. We observed that out of the six datasets, four of them show the successful attack rate as 1, while the other two have small rates as 0.01 and 0.03. For those datasets that have the successful attack rate as 1, there is a large portion of values in $\cal{B}$ that have a unique set of values in $\cal{A}$ to associate with to satisfy frequency equality {\bf???}. 

\nop{
\begin{figure}[ht]
\begin{center}
\begin{small}
\begin{tabular}{| c | c | c |}
\hline
Domain Size & Number of Distinct Values & rate \\ \hline
$100$ & $20$ & $0.04515$ \\ \hline
$100$ & $10$ & $0.8142$  \\ \hline
$1000$ & $20$ & $0.56277$ \\ \hline
$1000$ & $10$ & $1$ \\
\hline
\end{tabular}
\caption{Successful attack rate}
\label{fig:attack}
\end{small}
\end{center}
\end{figure}

{\bf Is it related to domain size? or the frequency distribution?} In contrast, in $D_2$ and $D_3$, at most one value in $\cal{B}$ has a unique set of values in $\cal{A}$ to associate with.
{\bf BOXIANG, THIS IS FAR FROM SATISIFCATION. PLS ADD THE FOLLOWING DETAILS:
(1) what's the successful rate for the 6 datasets? (2) what's the property of those datasets that give high successful rate?}
}

\nop{

\begin{figure}[ht]
\begin{center}
\begin{small}
\begin{tabular}{| c | c | c | c | c | c |}
\hline
$D_1$ & $D_2$ & $D_3$ & $D_4$ & $D_5$ & $D_6$ \\ \hline
1 & 0.0293 & 0.0131 & 1 & 1 & 1 \\ 
\hline
\end{tabular}
\caption{Successful attack rate}
\label{fig:rate}
\end{small}
\end{center}
\end{figure}
}

\vspace{-0.05in}
\section{Related Work}
\label{sc:related}

Data security is taken as a primary challenge introduced by the database-as-a-service ($DaS$) paradigm. 
To protect the sensitive data from the $DaS$ service provider, the client may transform her data so that the server cannot read the actual content of the data outsourced to it. 
A straightforward solution to data transformation is to encrypt the data while keeping the decryption key at the client side. 
Hacigumus et al. \cite{hacigumus2002providing} is one of the pioneering work that explores the data encryption for the $DaS$ paradigm. They propose an infrastructure to guarantee the security of stored data. Different granularity of data to be encrypted, such as row level, field level and page level, is compared. Chen et al. \cite{hacigumucs2002executing} develop a framework for query execution over encrypted data in the $DaS$ paradigm. In this framework, the domain of values of each attribute is partitioned into some bucket. The bucket ID which refers to the partition to which the plain value belongs serves as an index of ciphertext values. Both encryption methods are vulnerable against the frequency analysis attack as they only consider one-to-one substitution encryption scheme. Curino et al. \cite{curino2011relational} propose a $DaS$ system that provides security protection. Many cryptographic techniques like randomized encryption, order-preserving encryption and homomorphic encryption are applied to provide adjustable security. CryptDB \cite{popa2012cryptdb} supports processing queries on encrypted data. It employs multiple encryption functions and encrypts each data item under various sequences of encryption functions in an onion approach. 
Alternatively, Cipherbase \cite{arasu2013orthogonal} exploits the trusted hardware (secure co-processors) to process queries on encrypted data. These cryptographic techniques are not FD-preserving. 

Data encryption for outsourcing also arises the challenge of how to perform computations over the encrypted data. A number of privacy-preserving cryptographic protocols are developed for specific applications. For example, searchable encryption \cite{song2000practical,boneh2004public} allows to conduct keyword searches on the encrypted data, without revealing any additional information. However, searchable encryption does not preserve FDs. Homomorphic encryption \cite{smart2010fully} enables the service provider to perform meaningful computations on the data, even though it is encrypted. It provides general 
privacy protection in theory, but it is not yet efficient
enough for practice \cite{naehrig2011can}. 

Integrity constraints such as FDs are widely used for data cleaning. There have been very few efforts on finding data integrity constraints and cleaning of inconsistent data in private settings. Talukder et al. \cite{talukder2011detecting} consider a scenario where 
one party owns the private data quality 
rules (e.g., $FDs)$ and the other party owns the private data. These two parties wish to cooperate by checking the quality of the data in one party's database
with the rules discovered in the other party's database. They require that both the data and the quality rules need to
remain private. They propose a cryptographic approach for FD-based inconsistency detection in private databases without the use of a third party. The quadratic algorithms in the protocol may incur high cost on large datasets \cite{talukder2011detecting}. Barone et al. \cite{barone-qdb09} design a privacy-preserving data quality assessment that embeds data and domain look-up table values with Borugain Embedding. The protocol requires a third party to verify the (encrypted) data against the (encrypted) look-up table values for data inconsistency detection. Both work assume that the data quality rules such as $FDs$ are pre-defined. None of these work can be directly applied to our setting due to different problem definition and possibly high computational cost. 

Efficient discovery of FDs in relations is a well-known challenge in database research. Several approaches (e.g., TANE \cite{huhtala1999tane}, FD\_MINE \cite{yao2002fd_mine}, and FUN \cite{novelli2001fun}) have been proposed. \cite{papenbrock2015functional} classifies and compares seven FD discovery algorithms in the literature. 
 \cite{liu2010discover} presents an excellent survey of FD discovery algorithms.

\vspace{-0.1in}
\section{Conclusion and Discussion}
\label{sc:conclu}
In this paper, we presented $F^2$ algorithm that is FD-preserving and frequency-hiding. It can provide provable security guarantee against the frequency analysis attack, even under the Kerckhoffs's principle. Our experiment results demonstrate the efficiency of our approach. 

We acknowledge that $F^2$ does not support efficient data updates, since it has to apply splitting and scaling (Step 2.2) from scratch if there is any data update. For the future work, we will consider how to address this important issue. Another  interesting direction is to extend to malicious attackers that may not follow the outsourcing 
protocol and thus cheats on the data dependency discovery results. The problem of verifying whether the returned FDs are correct is challenging, given the fact that the data owner is not aware of any $FD$ in the original dataset. 

\section{Acknowledgement}
This material is based upon work supported by the National Science
Foundation under Grant SaTC-1350324 and SaTC-1464800. Any opinions,
findings, and conclusions or recommendations expressed in this
material are those of the author(s) and do not necessarily reflect the
views of the National Science Foundation.

%\vspace{-0.12in}
\bibliographystyle{abbrv}
{
\bibliography{bib,bib_pprl}  
}

\end{document}